\documentclass{gh}
\usepackage[super,compress]{cite}
\usepackage{braket}
\usepackage{graphicx}
\usepackage{cases}
\usepackage{afterpage}
\usepackage{gensymb}
\begin{document}

\markboth{Richard Herrmann}{Fractional  Schr\"odinger equation }

\catchline{}{}{}{}{}

\title{Solutions of the fractional Schr\"odinger equation via diagonalization -  A plea for the harmonic oscillator basis\\  part 1: the one dimensional case}
\author{\footnotesize Richard Herrmann}
\address{GigaHedron, Berliner Ring 80, D-63303 Dreieich \\
email:herrmann@gigahedron.com}

\maketitle

\begin{history}
\received{Day Month Year}
\revised{Day Month Year}
\end{history}

\begin{abstract}
A covariant non-local extention if the stationary Schr\"odinger equation is presented and it's solution in terms of Heisenbergs's matrix quantum mechanics is proposed.
For the special case of the Riesz fractional derivative, the calculation of corresponding matrix elements for the non-local kinetic energy term is performed fully analytically in the harmonic oscillator basis and leads to a new interpretation of non local operators in terms of generalized Glauber states.

As a first application, for the fractional harmonic oscillator the potential energy matrix elements are calculated and the  and the corresponding Schr\"odinger equation is
diagonalized. For the special case of invariance of the non-local wave equation under Fourier-transforms a new symmetry is deduced, which may be interpreted as an
extension of the standard parity-symmetry.
\end{abstract}


\ccode{PACS numbers:21.10.Dr;21.60Ev;02.40.Ky}


\section{Introduction}
{{ 
     When Heisenberg returned from his spring vacations at Helgoland to G\"ottingen in 1925, he carried in his rucksack a  derivation of the quantum harmonic oscillator discrete level spectrum \cite{hei25} as a first formulation of the mechanics of quantum particles in terms of coordinate- $\hat{x}$ and 
momentum- $\hat{p}$ operators, formally given as (using Dirac's bra-ket notation)
\begin{eqnarray}
\label{crrhh02}
 \hat{x} \rightarrow  \braket{m|x|n} \\
 \hat{p}  \rightarrow  \braket{m|p|n} 
 \end{eqnarray}
\noindent
with the canonical commutation relation
\begin{equation}
\label{crre01}
[ \hat{p}, \hat{x} ]  = -i \hbar  
\end{equation}
They were immediately interpreted as matrix operators \cite{bor25, bor26} and as constituents of his matrix mechanics, yielding the discrete energy levels of the harmonic oscillator as eigenvalues of a diagonal 
matrix 
\begin{eqnarray}
\label{crres1}
 H  = \frac{1}{2}(\hat{p}^2 + \hat{x}^2)
\end{eqnarray}
thus emphasizing the particle aspect of quantum particles.

Already one year later, Schr\"odinger, motivated by his view on the particle-wave duality,  presented his famous wave equation \cite{sch26a}, where  wave functions $\psi$ are determined  fulfilling certain boundary conditions:
\begin{eqnarray}
\label{crres1}
 H \psi = (T+V) \psi = i \hbar \partial_t \psi
\end{eqnarray}
This wave equation describes the mechanics of quantum particles from an alternative point of view, emphasizing the wave aspect of quantum particles. In this case, operators are realized as a conjugated set of operators and derivative operator, e.g.  in coordinate representation
\begin{eqnarray}
\label{crre02}
 \hat{x} \rightarrow x \\
 \hat{p}  \rightarrow -i \hbar  \partial_x 
\end{eqnarray}
and momentum representation, respectively.
\begin{eqnarray}
\label{crre03}
 \hat{p} \rightarrow p \\
 \hat{x}  \rightarrow i \hbar  \partial_p 
\end{eqnarray}
This approach may be categorized as wave mechanics of quantum particles.

Although Schr\"odinger demonstrated in a series of seminal papers, that both views are convertible,  his approach was considered heuristic by the Bohr school. 

Since Feynman's path integral formulation of quantum mechanics may be interpreted as a derivation of the  Schr\"odinger equation from first principles, the wave mechanic view based on the investigation of a corresponding wave equation is considered as a much more intuitive entry to quantum mechanics and therefore is the first choice in most textbooks on this subject \cite{sif49, sakurai, messiah, gre01b}.

In this chapter we will return to Heisenberg's matrix mechanics and extend his approach to the fractional case. We will collect arguments, why a diagonalization of these matrices especially on a harmonic oscillator basis is the appropriate choice to 
minimize the efforts to solve the fractional Schr\"odinger equation, to guarantee a correct treatment of singular long range kernels involved, to investigate the symmetries of the fractional quantum harmonic oscillator  and to obtain high precision results. 
}}

\section{From local to nonlocal operators}
The extension of standard quantum mechanics to the nonlocal case is a two step process with regard on spatial 
nonlocality.
First, the potential free Schr\"odinger equation, which means, the kinetic energy term $T$ alone  is extended to the nonlocal case and
then subsequently, the potential term  $V$ is minimally coupled conserving  local gauge invariance.

Let us illustrate this procedure for the special case of a nonlocal extension of the Sch\"odinger equation based on fractional calculus.  

 From a historical point of view fractional calculus may be considered as an extension of the concept of a 
 derivative operator from the integer order $n$ to arbitrary order $\alpha$, where $\alpha$ is an arbitrary
 real or complex value on R.
\begin{eqnarray}
\label{crreddd01}
\partial_x^n \rightarrow \partial_x^\alpha  
\end{eqnarray}
and therefore the kinetic term $T$ of the wave equation is extended using
\begin{eqnarray}
\label{crreddd02}
\hat{p}^2  &\rightarrow& \hat{p}^{2 \alpha}\\
 \hat{T}   &\rightarrow& \frac{1}{D^\alpha}\hat{T}^\alpha 
\end{eqnarray}
where $D^\alpha$ accounts for the unit conservation of the kinetic energy for the fractional case. 

Then, in a second step,  the potential energy term is added, which is unaffected by the nonlocalization procedure.
The space-fractional Schr\"odinger equation finally reads:
\begin{eqnarray}
\label{crres1}
 H \psi &=& ( \frac{1}{D^\alpha}\hat{T}^\alpha +V) \psi = i \hbar \partial_t \psi
\end{eqnarray}
This means, that the transition from local to nonlocal Schr\"odinger equation extends the role of the kinetic energy term from a purely geometry based static quantity to a  more dynamic element, while in standard quantum mechanics  the variety of different quantum phenomena is modeled by modifying the potential energy term only.

It is remarkable, that this view has influenced to date the way, fractional quantum mechanical problems have been treated analytically as well as numerically. Especially the proposed numerical treatment of the fractional Schr\"odinger equation  emphasizes the role of strategies and methods, which extend the classical approaches for a solution of a partial
differential equation 
to the fractional case, treating the fractional Laplace operator and the potential simultaneously. Within this category of solutions we have the finite difference methods  \cite{liz12, pop13, tar16, din17},  Galerkin finite element approximation \cite{zha10}  

The intention of this presentation is quite simple in that sense.
We will derive the matrix elements of the fractional kinetic energy term
\begin{eqnarray}
\label{crreddd02}
\hat{T}^\alpha = 
 \braket{m|\hat{p}^{2 \alpha}|n}
\end{eqnarray}
in the harmonic oscillator basis $\ket{n} $.

First applications of fractional calculus indeed were realized  on one-dimensional Cartesian space;  a problem was stated using a single coordinate e.g. 
a path-length $s$  as in Abel's treatment of the tautochrone-problem \cite{abel}, a time $t$  coordinate as in the description of
anomalous diffusion processes as exemplified e.g. by Metzler and Klafter \cite{met00} or in causal elastic wave equations \cite{nas13}.
In order to obtain a covariant definition of a fractional operator in multi-dimensional space, we first embed fractional calculus into a wider class of nonlocal
operators and start with a covariant definition of a general nonlocal operator in  Riemannian space and in a next step we 
consider the specific requirements for an application of fractional operators.

\section{Covariant definition of nonlocal operators}
We define the nonlocal pendant $\tilde{\mathcal{O}}(w)$ of a given local operator $\tilde{\mathcal{L}}$ as a product of two operations: 
First, the local operator is applied, then a nonlocalization operator $\tilde{\mathcal{G}}(w)$, realized as an integral operator with weight 
or kernel $w$, respectively, acts on the initially local result. 
\begin{figure}[t]
\begin{center}
\includegraphics[width=\textwidth]{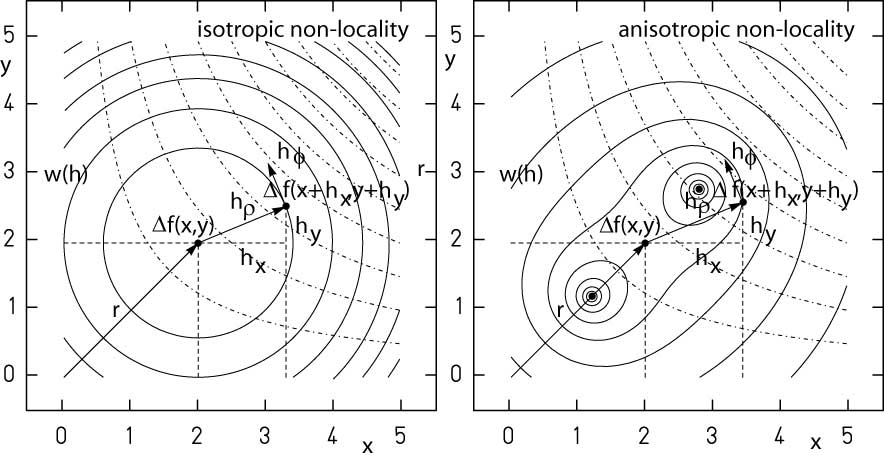}    
\caption{ Illustration of the  two step procedure for a generalized Laplace operator. First, the local part of the Laplace operator applied to a function $f$ yields $\tilde{\mathcal{L}}f = \Delta f$, which is       
plotted with dashed-dotted lines as $\Delta f$. 
The second step is the nonlocalization, realized as a weighted integral over the full region.
Two classes of weight functions $w(h)$ are shown with solid lines: On the left an isotropic weight contour, which is characteristic for e.g. the Gauss-normalizable,  Riesz- and tempered Riesz fractional singular weights respectively. On the right, a non-isotropic weight contour is shown, which is characteristic for e.g.
Erdely-Kober type weights. In multidimensional space, two special coordinate systems are useful to determine the distance $h$:  $h$ may be given in Carthesian- $\{h_x,h_y\}$ or, preferable for isotropic weights, in polar coordinates $\{h_r,h_\phi\}$. Since the presented procedure is covariant, the result is independent from a specific choice of 
coordinate sets.  
}
\label{fig1}
\end{center}
\end{figure}
\begin{equation}
\label{chxx01}
\tilde{\mathcal{O}}(w(|h|), \cdot)  =  
\tilde{\mathcal{G}}(w(|h|)) \otimes
\tilde{\mathcal{L}}(\cdot) 
\end{equation}
The nonlocalization $\tilde{\mathcal{O}}$ is a dual operator on $R^n \otimes R^n$, where two operations $\tilde{\mathcal{L}}$, $\tilde{\mathcal{G}}$ are given
on two different Riemann spaces. For a covariant realization of the nonlocalization $\tilde{\mathcal{O}}$
 each of  $\tilde{\mathcal{L}}$ and $\tilde{\mathcal{G}}$ must be given in covariant form. 

Let us express the requirement of covariance as form invariance under coordinate transforms, which on $R^n$  leaves the line element $ds^2$ invariant:
\begin{equation}
ds^2 = g_{\mu \nu}dx^\mu dx^\nu  \qquad \mu,\nu = 1, ..., n  
\end{equation}
This is a basic requirement to derive results, which at least are independent from a specific choice of a coordinate system.  In this way, we are directly lead  to the following specifications for the operators in (\ref{chxx01}).

First, the local operator $\tilde{\mathcal{L}}(\cdot)$ should transform as a tensor of a given rank $m$.
Classical examples are:

Gradient of a scalar field
\begin{eqnarray}
\label{chxxgrad}
\nabla^{i} \Psi =  g^{ij}\partial_{j}  \Psi  
\end{eqnarray}
divergence of a vector field, used e.g. in convection-diffusion equations 
\begin{eqnarray}
\label{chxxdiv}
& = & \nabla_{i}\Psi^{j} \\
&=& \partial_i \Psi^j +  
\left\{\begin{array}{c}
j     \\ ik           \\
\end{array} \right\}
 \Psi^k      
\end{eqnarray}
or, as a combination of divergence and gradient, the Laplace-operator contracted to a tensor of rank 0 (scalar) used in wave-equations:
\begin{eqnarray}
\label{chxxLaplace}
\bigtriangleup    \Psi     & = & \nabla^{j} \nabla_{j} \Psi \\
&=& g^{ij} \nabla_{i} \nabla_{j} \Psi 
      \qquad\qquad i,j=1,...,N                                      \\
   & = & g^{ij}(\partial_{i} \partial_{j} -
\left\{\begin{array}{c}
k      \\ ij           \\
\end{array} \right\}
                 \partial_{k})    \Psi       \\ 
  & = &\frac{1}{\sqrt{g}} \partial_{i}\,
        g^{ij} \sqrt{g} \, \partial_{j} \Psi 
                                   \end{eqnarray}
where
$\left\{\begin{array}{c}
k      \\ ij           \\
\end{array} \right\}$ is the Christoffel symbol
\begin{equation}
\left\{\begin{array}{c}
k      \\ ij           \\
\end{array} \right\}
=\frac{1}{2} g^{k l}(\partial_i g_{j l}+\partial_j g_{i l}-\partial_l g_{i j})
\end{equation}
\index{Christoffel symbol}
with $g$ being  the determinant of the metric tensor, $g= \det g_{ij}$
and
$\nabla_{n}$ is the covariant derivative.

Second, we postulate, that the fractional extension of a given local operator should not change the rank of the local operator. The fractional extension of a wave-equation should remain  a wave-equation with corresponding tensor characteristics, the fractional extension of the Fokker-Planck-equation should exhibit the same tensor properties as the local one.

Therefore  the nonlocalization operator  $\tilde{\mathcal{G}}(w(|h|))$ may only transform as a tensor of rank 0 (scalar). This is realized  as long as the weight is a function of the line element only.
\begin{equation}
w(|h|) \equiv  \frac{1}{N} w(|ds|)
\end{equation}
where the normalization factor $1/N$ accounts for correct dimensional treatment of the weight.

Therefore,  the covariant form of the nonlocalization operator is given as:
\begin{equation}
\label{chxxnonR}
\tilde{\mathcal{G}}_R(x^i) =\frac{1}{N}  \int_{R^n} \sqrt{|g(x)|} \, d^nx \, w(ds(x^i, x^j)) \times 
\end{equation}
with the covariant volume element $\sqrt{|g(x)|} \, d^nx$, acting on a standard local operator  $\tilde{\mathcal{L}}(x^j)$.
It should be emphasized, that the nonlocalization operator and the local operator may be appropriately defined in different 
coordinate systems, e.g. the local operator may be defined using carthesian coordinates, while the nonlocalization operator
is given in spherical coordinates.

Important cases are: 

\begin{itemize}
\item In $R^1$, using Cartesian coordinates $\tilde{\mathcal{G}}_R(h)$ is explicitly given by:
\begin{equation}
\label{chxxnonR1}
\tilde{\mathcal{G}}_R^1(x') =\frac{1}{N}  \int_\infty^\infty  dx' w(x,x') \times 
\end{equation}
acting on a local operator $\tilde{\mathcal{L}}(x)$.

\item In $R^2$, the use of polar coordinates $\{r, \phi\}$ is a good choice for the nonlocality operator because of the homogeneity requirement. This leads to the nonlocalization operator $\tilde{\mathcal{G}}_R(r',\phi')$
\begin{equation}
\tilde{\mathcal{G}}_R(r',\phi') =\frac{1}{N}  \int_{0}^{\infty} r' dr' \int _0^{2 \pi} d \phi'  w(r', \phi', x^j)  \times 
\end{equation}
acting on a local $\tilde{\mathcal{L}}(x^j)$.

\item In $R^3$, the use of spherical coordinates $\{r, \phi, \theta' \}$ yields for  the nonlocalization operator $\tilde{\mathcal{G}}_R(r',\phi', \theta')$
\begin{equation}
\tilde{\mathcal{G}}_R(r',\phi', \theta') =\frac{1}{N}  \int_{0}^{\infty} r'^2 dr' 
\int _0^{2 \pi} d \phi'
\int _0^{\pi} \sin(\theta') d \theta'
    w(r', \phi', \theta' x^j)  \times 
\end{equation}
acting on a local $\tilde{\mathcal{L}}(x^j)$.
\end{itemize}

    
An explicit realization of a valid nonlocalization  may be given in terms of sets of classical weights, which are used e.g. as generating sets of generalized  functions like Dirac's $\delta$-function. The most famous representative of this function class is the Gaussian:
\begin{equation}
\label{www1Gauss}
w_{Gauss}(|h|, \sigma) = \frac{1}{N_{Gauss}(n)} e^{-h^2/\sigma^2}
\end{equation}
with the norm
\begin{equation}
\label{chxxnonNormGauss}
N_{Gauss}(n) = \pi^{n/2} \sigma^n
\end{equation}
where $n$ is the space dimension.
The smooth transition from nonlocal to local limit of $\tilde{\mathcal{G}}_R(x^i)$ results as any  
realization of a limiting procedure, for the Gaussian as $\lim_{\sigma \rightarrow 0}$.


Therefore we have defined a procedure for realizing the transition from local to generalized nonlocal operators, which guarantees covariant results, as long as the nonlocalization operator $\tilde{\mathcal{G}}$, the weight $w$ and the local operator$\tilde{\mathcal{G}}$ are defined appropriately.

In the following we will investigate a special case of a nonlocal operator,  the fractional time-independent  Schr\"odinger equation, which is invariant under Galilei-transformations. 
Restricting the nonlocalization procedure to the spatial part  and
keeping the time coordinate $t$ local, allows for a classical treatment of the potential term $V$, which means minimal
coupling of the zeroth component of the electromagnetic field $A_\mu = \{V, A_i\}$ conserves local gauge invariance for the fractional case, too.  

The covariant local stationary, potential free Hamilton operator $\tilde{\mathcal{H}}_{free}$ reduces to the
kinetic energy term $T$:  
\begin{eqnarray}
\label{chxxSGL0}
\tilde{\mathcal{H}}_{free} &\equiv&\hat{T} \\
     &=&( -\frac{\hbar^2}{2 m}
  \frac{1}{\sqrt{g}} \partial_{i}\,
        g^{ij} \sqrt{g} \, \partial_{j}) \times
\end{eqnarray}
and the nonlocal generalized covariant Schr\"odinger equation with potential $V$ results as
\begin{eqnarray}
\label{chxxSGL1}
(\hat{T}^\alpha + \hat{V} - E) \psi &=& 0
\end{eqnarray}
and  is explicitly given by:
\begin{eqnarray}
\label{chxxSGL2}
(-\tilde{\mathcal{G}} \times
  \frac{\hbar^2}{2 m}
\frac{1}{\sqrt{g}} \partial_{i}\,
        g^{ij} \sqrt{g} \, \partial_{j} + V)  \psi = E \psi
\end{eqnarray}
This is the starting point for a covariant  fractional quantum mechanics producing results, which are invariant under all coordinate transformations, which leave the line element invariant, see e.g. Moon and Spencer for a collection \cite{moo88}.  

To obtain the eigenvalue spectrum and eigenfunctions for a given potential we propose a diagonalization of the corresponding matrix representation. The main advantage of this approach is a separated treatment of the kinetic and
potential energy terms. Once the matrix elements are calculated for the fractional kinetic energy term  $\hat{T}^\alpha $ within
a given basis $\ket{n} $, these matrix elements remain unchanged when the potential energy term is altered, because only these 
matrix elements need to be recalculated. 

Since the  matrix elements $ \braket{m|V|n}$ for the potential energy are not affected by the nonlocal generalization of the kinetic energy term, these matrix elements can be adopted without modification from the local case.

We propose as an appropriate basis  $\ket{n}$ the harmonic oscillator wave functions. These are well behaved
on the full domain $R^n$, which makes them ideal candidates, since an infinitely spaced nonlocalization operator    $\tilde{\mathcal{G}}$ is involved. Furthermore, the weak singularities, which  are essential for the fractional calculus are
treated without significant problems.  This is in contrast to direct methods, which react very sensitive in the vicinity of singularities.

Furthermore, the matrix representation of the kinetic and potential terms is very helpful to deduce inherent symmetries.
As a first application, we will investigate the relationship between the fractional kinetic energy and the potential 
energy matrix elements for the fractional harmonic oscillator, which will reveal the very close relationship between these
two energy terms as a result of the equivalence of a coordinate- or momentum representation and as a consequence we obtain
a very condensed, efficient explicit calculation rule of the matrix elements.

In order to calculate the matrix elements of the fractional kinetic energy matrix elements and apply these results to calculate numerically
first results for the harmonic oscillator, in the next section we will 
consider the role of singularities and long-range interactions in fractional calculus and investigate  their influence
on the eigenvalue spectrum of the fractional kinetic energy operator.

\section{Covariant nonlocalization operators}
The covariant nonlocalization operator $\tilde{\mathcal{G}}$ (\ref{chxxnonR}) is realized as 
an integral of convolution type with an at first arbitrarily chosen kernel. 
In this section, we want to introduce two different classes of nonlocal kernels, which represent two
different classes of kernels.

The first class is based on Dirac's definition of the $\delta$-function, where kernels are characterized 
by non-singular kernels and the second class is
based on  fractional calculus, where kernels are weakly singular and of long range type.

We will present the main properties of these different weight-families and will investigate especially, which fractional derivative kernels may be  used as a basis for more general weight definitions for nonlocalization operators. 

For reasons of simplicity we will restrict our presentation to the one dimensional case $R^1$.

First, following Dirac's approach, we may construct a transition from nonlocal to local description with a set of good (in the sense of Lighthill \cite{light})  functions $w$ (non-singular kernels), which 
in the limit $n \rightarrow \infty$ yields Dirac's $\delta$-function  
\begin{eqnarray}  
\label{nl_dirac}
 f(x) 
 &=& 
 \lim_{n \rightarrow \infty }\int_{-\infty}^{\infty} dx'  w(x-x', n) f(x')  
 = 
\int_{-\infty}^{\infty} dx'  \delta(x-x')  f(x') \nonumber \\ 
\end{eqnarray} 
with normalizable weights:
\begin{eqnarray}  
\label{nl_dirac_norm}
\frac{1}{N(n)} \int_{-\infty}^{\infty} dx'  w(x',n)  &=& 1  
\end{eqnarray}
A classic example for a valid weight is the Gaussian (\ref{www1Gauss}).
The transition from $n \in N$ to $n \in R$ simply results in a smooth parametrization of the Gaussian width $\sigma$.
For finite $n$ nonlocality enters via an averaged, normalized distribution of function values with a non-vanishing weight
for different positions. Applied to a differential operator e.g. the derivative consequently results in an spatial 
weighted average of the same operator.

On the other hand we have weakly singular kernels, historically arising from Cauchy's concept of a derivative definition, restricted to the real axis:   
\begin{eqnarray}  
\label{nl_cauchy0}
 f^{(n)}(x) 
 &=& 
 \frac{n!}{2 \pi i}  \oint dx'  \frac{f(x')}{(x-x')^{1+n}  } \quad n \in N
\end{eqnarray}
where  $f^{(n)}$ denotes the n-th derivative with respect to $x$. It should be emphasized, that already for any $n \in N$ this definition gathers weighted information of the function $f$ on full domain $R^1$.  

The canonical transition from integer $n \in N$ to
real $\alpha  \in R$ leads to a continuous definition of a derivative of function $f$ as an integral over the full $R^1$: 
\begin{eqnarray}  
\label{nl_cauchya}
 f^{(\alpha)}(x) 
 &=& 
  \frac{\Gamma(1+\alpha)}{2 \pi i }  \oint dx'  \frac{f(x')}{(x-x')^{1+\alpha}  } \quad  \beta \in R
\end{eqnarray}
So, while in coordinate space we have a smooth transition of $x^\alpha$ values,
 the corresponding $p^{\alpha}$ values in momentum space are given by the non local operation  (\ref{nl_cauchy}).

It should be emphasized, that we do not intend to derive a fractional derivative directly: We are exclusively interested 
in finding a corresponding nonlocalization operator. To be clear we introduce a new variable $\beta$, defined as a shift in $\alpha$ values.     
\begin{eqnarray}  
\label{nl_beta}
 \beta &=& \alpha-1
\end{eqnarray}
 Consequently (\ref{nl_cauchya}) takes the form  
\begin{eqnarray}  
\label{nl_cauchy}
 f^{(\beta-1)}(x) 
 &=& 
  \frac{\Gamma(\beta)}{2 \pi i }  \oint dx'  \frac{f(x')}{(x-x')^{\beta}  } \quad  \beta \in R
\end{eqnarray}
Restricting to the real axis, operator (\ref{nl_cauchy})  is equivalent to the Liouville definition of a left/right fractional derivative.
 \begin{eqnarray}  
\label{nl_gtestLL}
 f^{(\beta-1)}(x)  &=&  L^+ f(x) = \frac{\Gamma(\beta)}{\pi}  \int_{0}^{\infty} dx'  \frac{f(x+x')}{x'^{\beta}}  \\
 \label{nl_gtestLR}
 f^{(\beta-1)}(x)  &=&  L^- f(x) =\frac{\Gamma(\beta)}{\pi}  \int_{0}^{\infty} dx'  \frac{f(x-x')}{x'^{\beta}}   
\end{eqnarray} 
In the following,  we want to define a nonlocalization operator $\tilde{\mathcal{G}}$, 
 acting on a function $g$, which already is a result of a  local operation $\tilde{\mathcal{L}}$ (e.g. a derivative operator) $\tilde{\mathcal{L}} f = g$.
 Minimum requirements  for  $\tilde{\mathcal{G}}$ are parity conservation
 \begin{eqnarray}  
\label{nl_gtest1}
 \Pi (\tilde{\mathcal{G}} g) &=& \Pi g = \Pi  (\tilde{\mathcal{L}} f) 
\end{eqnarray}
where $\Pi$ is the parity operator. 
Consequently, a sign-shift in $x'$ is not tolerable.

The Liouville operators $L^+$ (\ref{nl_gtestLL}) and  $L^-$ (\ref{nl_gtestLR})
do not fulfill this condition. This may easily be shown applying e.g.  
the Liouville operator  $L^+$   on a set of  test functions with good parity, e.g. the trigonometric functions:
 \begin{eqnarray}  
\label{nl_gtest2}
 L^+ f(x) = \frac{\Gamma(\beta)}{\pi}  \int_{0}^{\infty} dx'  \frac{1}{x'^{\beta}} \cos(k (x+x'))&=&
 k^{\beta-1}\sin(k x + \beta \pi/2) \nonumber\\ & & \\
 L^+ f(x) = \frac{\Gamma(\beta)}{\pi}  \int_{0}^{\infty} dx'  \frac{1}{x'^{\beta}} \sin(k (x+x'))&=&
 k^{\beta-1}\cos(k x + \beta \pi/2) \nonumber \\
 \end{eqnarray}
Obviously parity is not conserved.

To solve this problem, we compensate the sign-change by applying Moivre's formula, which allows to combine the
areas of integration for $L^+$ and  $L^-$:  
\begin{eqnarray}  
\label{nl_cauchyfracteven}
 f^{(\beta-1)}(x) 
 &=& 
 \frac{\Gamma(\beta)}{\pi} \cos(\frac{\beta \pi}{2}) \int_{-\infty}^{\infty} dx'   \frac{f(x')}{(x-x')^{\beta} } \quad \textrm{parity even}\\
 \label{nl_cauchyfractodd}
 f^{(\beta-1)}(x) 
 &=& 
  \frac{\Gamma(\beta)}{\pi} \sin(\frac{\beta \pi}{2}) \int_{-\infty}^{\infty} dx'   \frac{f(x')}{(x-x')^{\beta} }\quad \textrm{parity odd}
\end{eqnarray}
Hence there are two possible extensions of the Cauchy-integral from integer $n$ to the fractional $\beta$ case, which yields
two different definitions of a fractional derivative. 

The first one (\ref{nl_cauchyfracteven}) is parity conserving and coincides with
even derivatives in the limit $\beta \rightarrow 2n+1$. This we call Riesz type \cite{riesz}, because the Riesz-weight  $w_{Riesz}$ (\ref{chxxnonRiesz0}), which is  thus derived,  is the most
prominent representation of fractional even weights. 

The second one (\ref{nl_cauchyfractodd}) has odd parity and therefore parity is not conserved. It coincides with
odd derivatives in the limit $\beta \rightarrow 2n$. This we call Feller type \cite{feller} , because the Feller-weight, which is  thus derived, is the most
prominent representation of fractional odd weights.

In the following, we will proceed to investigate the properties of Riesz- type kernels, since
the Riesz-weight fulfills the requirements of isotropy and homogeneity on $R^n$: 
\begin{equation}
\label{chxxnonRiesz0}
w_{Riesz}(h, \beta) = \frac{1}{N_{Riesz}} |h|^{-\beta} 
\end{equation}
The normalization of the
  Riesz-weight is given on $R^n$ by
\begin{equation}
\label{chxxnonNormNN}
N_{Riesz}(n) = \frac{\pi^{1+n/2}}
{2^\beta  \Gamma(1+\beta/2) \Gamma((\beta+n)/2) \cos(\beta \pi/2) }
\end{equation}
which reduces in the one-dimensional case to
\begin{equation}
\label{chxxnonNormNN1}
N_{Riesz}(1) = \frac{\pi} {\Gamma(\beta) \cos(\frac{\beta \pi}{2})}
\end{equation}
as already derived  in (\ref{nl_cauchyfracteven}).

Since we consider parity conservation as a basic requirement for a general nonlocalization operator, we use function  (\ref{nl_cauchyfracteven}) of the above introduced two possible definitions as a basic example for a valid general weight for the nonlocalization, or in 
other words: the weights we investigate in the following are all even weights conserving parity.    

Thus we may consider as a nonlocalization operator $\tilde{\mathcal{G}}$ 
\begin{equation}  
\label{nl_cauchyfractevenR}
\tilde{\mathcal{G}}(w_{even}) f(x) = 
 \int_{-\infty}^{\infty} dh   w_{even}  f(x+h)  \quad  \textrm{parity even}
\end{equation}
with weights of Riesz-type.

With this last step  we have found the correct expression for a possible multiplicative nonlocalization operator $\tilde{\mathcal{G}}$.  
In the simple case of using a Riesz weight, $\tilde{\mathcal{G}}(w_{Riesz})$ is equivalent to the Riesz-derivative. The interpretation is different in terms of a 
limit of $\beta$, though.

The Riesz-derivative operator  coincides with the standard
Laplacian $-\Delta$ in the limit $\beta \rightarrow 3$ (which corresponds to $\alpha \rightarrow 2$) and therefore is often abbreviated as $(-\Delta)^{\alpha/2}$.
But in contrast to the Riesz-derivative definition, for $\tilde{\mathcal{G}}$ we are interested in the limit $\beta \rightarrow 1$.

Hence we may apply the abbreviation 
\begin{eqnarray}
\label{chxxnonGGG}
\tilde{\mathcal{G}}(w_{Riesz}) &=& |-\Delta|^{\alpha/2} f(x)= |-\Delta|^{(\beta-1)/2}f(x) \\
&=&  \frac{\Gamma(\beta)}{\pi} \cos(\frac{\beta \pi}{2}) \int_{-\infty}^{\infty} dx'   \frac{f(x')}{(x-x')^{\beta} }\\ 
&=&  \frac{\Gamma(\beta)}{\pi} \cos(\frac{\beta \pi}{2}) \int_{-\infty}^{\infty} dh  |h|^{-\beta} f(x+h) 
\end{eqnarray}
for the Riesz-weight on $R^1$, which in the case $\beta \rightarrow 1$ reduces to the unit-operator.

The eigenfunctions of this non local operator are the trigonometric functions $\psi^-(x,k) = \sin(k x)$ and $\psi^+(x,k) =\cos(kx)$ and the eigenvalues are given by
\begin{equation}
\label{chxxnonGEigen}
\tilde{\mathcal{G}}(w_{Riesz}) \psi^\pm(x,k)= e_k(\beta) \psi^\pm(x,k) =  |k|^{\beta-1}\psi^\pm(x,k) 
\end{equation}
Note, that the parity of states is conserved and the eigenvalue spectrum tends to $e_k(\beta=1)  = +1$ for $\beta \rightarrow 1$. Furthermore, for $\beta \rightarrow 3$ the eigenvalue spectrum is given by $e_k(\beta=3)  = +k^2$, which is of course not
the second derivative of    $\psi^\pm(x,k)$.

It is now clear, why we introduced a redefinition of a fractional $\beta$ in terms of an $\alpha$-range shift, with our interpretation of $\beta$ is a  measure for the covariant nonlocalization $\tilde{\mathcal{G}}$.

It coincides with
the standard Riesz fractional derivative only when applied  to the standard local Laplace-operator.
In case of the fractional Schr\"odinger
equation results are therefore identical for both operators, as long as the Riesz-weight is used.  In case of other tensor equations our approach guarantees covariance, while usual fractional derivatives in general lack this property.
 
We have demonstrated, that a covariant nonlocalization of a given local operator is possible and may be
realized either using normalizable weights in analogy to Dirac's approach for a series definition of a generalized function
or using a fractional calculus approach with weakly singular kernels. 

In view of a numerical treatment special emphasis should be placed on an adequate treatment of singularities and
long range properties of the kernels applied. Hence, in the following section we will investigate the consequences of using
different kernel types for a covariant nonlocalization operator, especially for the nonlocal Laplace-operator, used
for the kinetic energy term in the nonlocal Schr\"odinger equation.

 \section{Specific remarks on the use of fractional calculus weights }
According to Gorenflo et. al \cite{gor14}, fractional calculus deals with integro-differential equations where the integrals are of convolution type and exhibit long range weakly singular kernels of power-law type \cite{gor97}. 
A numerical solution of the nonlocal fractional Schr\"odinger equation therefore has to incorporate the singularity at the origin as well as the long range interaction appropriately.

In the one dimensional case, the nonlocal stationary fractional Schr\"odinger equation is given according to (\ref{chxx01}) by:
\begin{eqnarray}
\label{chxxsglfree}
\tilde{\mathcal{H}}(w(|h|))  \psi &=&  
(\tilde{\mathcal{T}}(w(|h|)) + V)\psi = E \psi \\
&=&
(\tilde{\mathcal{G}}(w(|h|)) \otimes T + V)\psi = E \psi 
\end{eqnarray}
where $\tilde{\mathcal{H}}(w(|h|))$ is the fractional extension of the standard Hamiltonian, consisting of two terms, the fractional extension of the kinetic energy operator $\tilde{\mathcal{T}}(w(|h|) $ and the standard potential energy operator $V$.

It is important to emphasize, that the nonlocal Hamiltonian consists of a nonlocal kinetic and a local potential term. An analytic as well as a  numerical
treatment should take advantage of this fact. 

The matrix representation of quantum mechanics is an ideal starting point  for
a separate treatment of these two terms. The non local matrix elements for the kinetic part are calculated only once  and
may be used unchanged for many different potentials. This is in contrast to other methods, where
the complete Hamiltonian is involved, e.g. classical shooting methods \cite{he13a}.

The fractional kinetic energy operator $\tilde{\mathcal{T}}(w(|h|) $ itself is the extension of the standard kinetic energy term by the nonlocalization operator $(\tilde{\mathcal{G}}(w(|h|))$
\begin{eqnarray}
\label{chxxsglfreeGG}
\tilde{\mathcal{T}}(w(|h|) \psi  &=& E \psi \\
&=&
\tilde{\mathcal{G}}(w(|h|)) \otimes T  \psi \\
&=&
\tilde{\mathcal{G}}(w(|h|)) \otimes ( -\frac{\hbar^2}{2 m} \partial_x^2)  \psi ,
 \end{eqnarray}
where the specific form of the nonlocalization depends on the weight kernel used.

\begin{table}[t]
\tbl{
Eigenvalue spectrum of the one dimensional normalized nonlocalization
operator $\tilde{\mathcal{G}}(w(|h|)$ (\ref{chxxsglfreeGGsolutions1})  using the Riesz- (\ref{chxxnonRiesz0}), tempered exponential- (\ref{chxxnonRiesz1}) and tempered $\cos$- (\ref{chxxnonRiesz2}) weight in the upper part of the table. In the lower part non-singular weights using the Gaussian (\ref{chxxnonGauss0})  and non-singular Riesz (\ref{chxxnonRiesz01}) are used. $K_n(x)$ is the Bessel-K function}
{\begin{tabular}{l|ll}
\hline\noalign{\smallskip}
weight & eigenvalue spectrum of nonlocalization operator  \cr
\noalign{\smallskip}
\hline\noalign{\smallskip}
$|h|^{-\beta}$ &  
$|k|^{\beta-1}$ & $0 \leq \beta \leq 1$
\cr
$|h|^{-\beta} e^{-b |h|}$ & 
$\frac{1}{2}(|b - i k|^{\beta-1} + |b+i k|^{\beta-1})/\sin(\frac{\beta \pi}{2})=$
& $0 \leq \beta \leq 1$      
\cr
&
$ (b^2 +k^2)^{(\beta-1)/2}
{_2}F_1 (\frac{\beta-1}{2}, -\frac{\beta-1}{2};\frac{1}{2}; \frac{k^2}{b^2+k^2})/\sin(\frac{\beta \pi}{2})$  
&, $b >0$    
\cr
$|h|^{-\beta} \cos(b h)$ &
$|b -  k|^{\beta-1} + |b+ k|^{\beta-1}$
& $0 \leq \beta \leq 1$     
 \cr
\hline\noalign{\smallskip}
$e^{-b^2 h^2}$ &
$e^{-\frac{1}{4} b^2 k^2}$
& $b>0 $     
 \cr
$(h^2 + b^2)^{-\beta/2}$ &
$|k|^{(\beta-1)/2} K_{\frac{\beta-1}{2}}( b |k|)$
& $0 < \beta, b>0 $     
 \cr
\end{tabular}}
\label{c113tab1}
\end{table}

The fractional weights, which we want to discuss in the following are the classical Riesz-weight
\begin{equation}
\label{chxxnonRiesz0}
w_{Riesz}(h, \beta) = \frac{1}{N_{Riesz}} |h|^{-\beta} 
\end{equation}
as well as the tempered exponential Riesz-operator \cite{sab15, gad13}
\begin{equation}
\label{chxxnonRiesz1}
w_{tRiesz}(h, \beta, b) = \frac{1}{N_{Riesz}} e^{-b|h|}|h|^{-\beta} 
\end{equation}
and the tempered cosine Riesz-operator
\begin{equation}
\label{chxxnonRiesz2}
w_{cRiesz}(h, \beta, b) = \frac{1}{N_{Riesz}} \cos(b h)|h|^{-\beta} 
\end{equation}
These fractional weights may be contrasted to typical non-singular weights, e.g. the Gaussian
\begin{equation}
\label{chxxnonGauss0}
w_{Gauss}(h, b) =  \frac{1}{N_{Gauss}} e^{-b^2 h^2} 
\end{equation}
as well as the non-singular shifted Riesz-operator 
\begin{equation}
\label{chxxnonRiesz01}
w_{sRiesz}(h, \beta, b) =  \frac{1}{N_{sRiesz}} (h^2 + b^2)^{-\beta/2} 
\end{equation}
All weights fulfill the requirements of isotropy and homogeneity on $R^n$. 

\begin{figure}[t]
\begin{center}
\includegraphics[width=\textwidth]{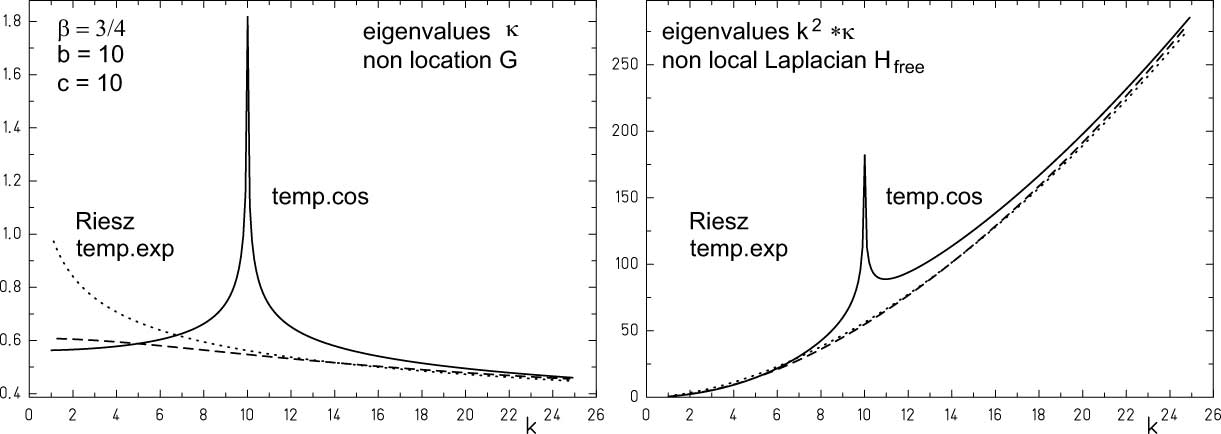}    
\caption{The energy spectra for the singular weights (\ref{chxxnonRiesz0})-(\ref{chxxnonRiesz2}) with parameters $\beta=3/4, b=10$. On the left the nonlocalization
spectrum $\kappa(k, w(|h|))$ (\ref{chxxsglfreeGGsolutions1}). On the right the complete nonlocal kinetic energy spectrum
$k^2 \kappa(k, w(|h|))$ of the nonlocal kinetic energy operator $\tilde{\mathcal{T}}(w(|h|) $ from (\ref{chxxsglfreeGG}),
which corresponds to the free fractional Hamiltonian $H_{free}$. Thick line indicates the Riesz weight, dashed line depicts
the tempered exponential weight and dotted line shows the tempered $\cos$ weight.
 }
\label{riesztau}
\end{center}
\end{figure}

The eigenvalue spectrum of the nonlocal kinetic energy term (\ref{chxxsglfreeGG}) may be easily determined by harmonic analysis, since
its eigenfunction are the trigonometric $\psi^+ (k) = \cos(k x)$ and $\psi^-(k) = \sin(k x)$-functions for the local and nonlocal contribution each.

We obtain
\begin{eqnarray}
\label{chxxsglfreeGGsolutions0}
  -\frac{\hbar^2}{2 m} \partial_x^2 \psi^\pm (k)  &=& \frac{\hbar^2}{2 m} k^2 \psi^\pm (k) \\
\label{chxxsglfreeGGsolutions1}
\tilde{\mathcal{G}}(w(|h|)) \psi^\pm (k) &=& \kappa(k, w(|h|)) \psi^\pm (k)
 \end{eqnarray} 
In table \ref{c113tab1} we have listed the analytic results for  $\kappa(k, w(|h|))$.

\begin{figure}[t]
\begin{center}
\includegraphics[width=\textwidth]{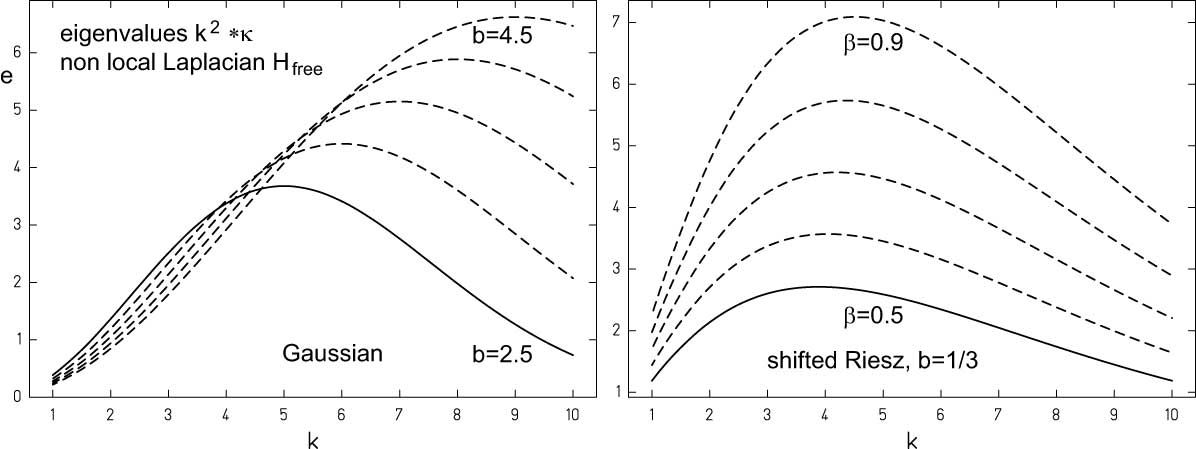}    
\caption{The complete not normalized nonlocal kinetic energy spectrum
$k^2 \kappa(k, w(|h|))$ of the nonlocal kinetic energy operator $\tilde{\mathcal{T}}(w(|h|) $ from (\ref{chxxsglfreeGG}),
which corresponds to the free fractional Hamiltonian $H_{free}$ for the non-singular Gaussian weight  (\ref{chxxnonGauss0}) on the left for different $b$, which in the limit for $b \rightarrow \infty$ leads to the parabolic local kinetic energy spectrum  $e = k^2$ and shifted Riesz weight (\ref{chxxnonRiesz01}) ($b=1/3$) on the right for different $\beta$, which in the limit for $b \rightarrow 0$ leads to the
fractional Riesz weight with the spectrum
 $e = k^{\beta+1}$. All spectra show a break down in the spectrum for large momenta
$k \gg 0$.
 }
\label{riesztau1}
\end{center}
\end{figure}

In figure~\ref{riesztau} we have plotted the resulting energy spectra for the singular weights (\ref{chxxnonRiesz0})-(\ref{chxxnonRiesz2}) and in figure~\ref{riesztau1} for  the non-singular weights 
(\ref{chxxnonGauss0})- (\ref{chxxnonRiesz01}). 

A comparison of the functional behavior of the analytically calculated energy spectra for different weights reveals
the consequences of using of weakly singular weights.

Let us  first discuss the influence of singularities in nonlocalization kernels on the calculation of the 
spectrum of the nonlocal Laplacian and the consequences of an appropriate treatment. 
According to Heisenberg's uncertainty relation the weak singularity for weights  (\ref{chxxnonRiesz0})-(\ref{chxxnonRiesz2}) at $h=0$
influences the eigenvalue spectrum for large momenta $k \gg 0$. From figure  \ref{riesztau} we may deduce, that the asymptotic
behavior of the eigenvalues of the nonlocalization operator $\tilde{\mathcal{G}}(w(|h|))$ is independent of a specific
choice for a weight. Differences only occur for small $k$-values. In this region, the nonlocal Laplacian is dominated by the
parabolic term $k^2$ of the local kinetic energy term and therefore, besides the cusp for the tempered cosine-weight the
overall spectrum for small and large k-values is similar for all singular weights investigated.  

When regarding non-singular weights a comparison with spectra
presented in figure  \ref{riesztau1} reveals a different
behavior. 

Independent from a specific choice of weight and its parameters there always exists  a critical $k$-value, above which the expected overall power
law for the spectrum  breaks down and we observe a beginning drop for all curves instead. This is the major difference when using non-singular weights.

The explanation is quite simple. Non-singular weights may be expanded in a Taylor-series at $h=0$. 
\begin{equation}
\label{chxxnonsing}
w_{non-singular}(h) = \sum_{i=0}^\infty c_{2 i}  h^{2 i}
\end{equation}
and therefore we have a smooth transition from e.g. cubic to quadratic to linear and finally constant behavior when the size of $h$ is step by step decreased.
\begin{equation}
\label{chxxnonsing1}
\lim_{h\rightarrow 0} w_{non-singular}(h) = c_0 = \textrm{const}
\end{equation}
For higher momenta the eigenfunction $\cos(k x)$ and $\sin(k x)$ respectively exhibit high frequence sign-oscillations  leading to canceling contributions for
the almost constant non-singular weight function, which yields in a breakdown of the power-law behavior of the energy spectrum.

For singular weights there is a kind of scale-invariance: Decreasing the value of $h$   does not change the dominant contribution
of the singularity at $h=0$. As a consequence the power law behavior of the eigenvalue spectrum of the nonlocal Laplacian is
conserved even in the limit $k \rightarrow \infty$.

From time to time the question arises, whether fractional calculus and fractals have more in common than only the name \cite{tat95, lei03, cal11a}. Within
the above discussed context the answer may be yes. At least, both research areas emphasize the role of self-similarity.

As a first result of our investigations we obtain a fundamental requirement for a successful numerical approach to the solution of the
fractional nonlocal Schr\"odinger equation: an appropriate treatment of the weak singularities is essential for the
asymptotic accuracy of the calculated eigenfunctions and -values.

\begin{figure}[t]
\begin{center}
\includegraphics[width=\textwidth]{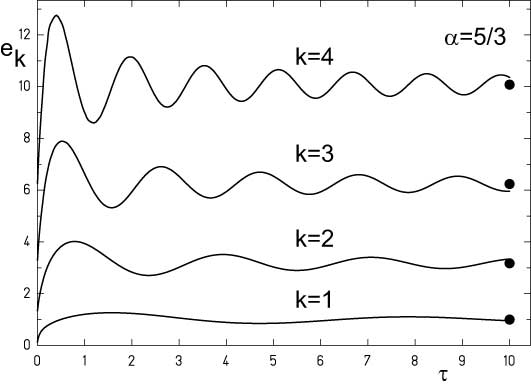}    
\caption{For $\beta=2/3$ the lowest four eigenvalues ($e_k$, $k,0,...,4$) of the Riesz fractional Laplace-operator are plotted for increasing finite integral bound $\tau$. Points indicate the energy values at $\tau = \infty$, which corresponds to the standard definition of the Riesz fractional derivative on the full domain. The finite cut-off results in oscillatory behavior of the eigenvalues.   
}
\label{riesztau2}
\end{center}
\end{figure}

However there is a second difficulty, which arises from the long-range behavior of the singular weights. The area of integration
is infinite, but in practical calculations sometimes only a finite area is considered. A classic example is the Schr\"odinger
equation with infinite well potentials, where some authors explicitly restrict the integration area to the finite size of the potential well \cite{guo06, don07, las10, bay12a, bay12b} which {\it{per se}} leads to wrong results. Similarly, the proposed
fractional extension of the WKB-approximation, used e.g. by Laskin \cite{laskin}  to give an approximate solution for the
fractional harmonic oscillator implicitly restricts the area of integration to the classically allowed region within the classic
turning points, which is a crude approximation already within the conceptual framework. 

In figure~\ref{riesztau2} we show the consequences for a finite limit of the long range contribution of the Riesz weight, introducing a cut off
$\tau>0$ and replacing the infinite integral bounds for the nonlocalization operator $\tilde{\mathcal{G}}(w_{Riesz})$ in (\ref{chxxnonGGG}) by
\begin{eqnarray}
\label{chxxnonGGGrestricted}
\tilde{\mathcal{G}}(w_{Riesz}, \tau) 
&=&  \frac{\Gamma(\beta)}{\pi} \cos(\frac{\beta \pi}{2}) \int_{-\tau}^{\tau} dx'   \frac{f(x')}{(x-x')^{\beta} }\\ 
&=&  \frac{\Gamma(\beta)}{\pi} \cos(\frac{\beta \pi}{2}) \int_{-\tau}^{\tau} dh  |h|^{-\beta} f(x+h) 
\end{eqnarray}
such that the eigenvalue spectrum of the nonlocal Laplacian now depends on the cut-off $\tau$ and is given by:
\begin{eqnarray}
\label{chxxnonGGGrestricted}
\tilde{\mathcal{L}}(w_{Riesz}, \tau) \psi^{\pm}(k)
&=&  -\frac{1}{\Gamma(1-\beta) \sin(\frac{\beta \pi}{2})} 2 k^2 \tau^{1-\beta} \times \nonumber \\
&&  {_1}F_2(\frac{1}{2}-\frac{\beta}{2};\frac{1}{2}, \frac{3}{2}-\frac{\beta}{2};-\frac{1}{4}k^2 \tau^2)  \psi^{\pm}(k)
\end{eqnarray}
which in the limit $\tau \rightarrow \infty$ reduces to the standard result:
\begin{eqnarray}
\label{chxxnonGGGrestrictedinf}
\lim_{\tau \rightarrow \infty } \tilde{\mathcal{L}}(w_{Riesz}, \tau) \psi^{\pm}(k)
&=&  k^2 |k|^{\beta-1}\psi^{\pm}(k)
\end{eqnarray}
In figure~\ref{riesztau2} the lowest eigenvalues of the nonlocal Laplacian are plotted as a function of the cut-off $\tau$.
We observe an oscillatory behavior of the calculated energies around the infinite cut-off result.

Deducing from this we demand as a second requirement for a consistent realization of a numerical approach an appropriate 
treatment of the long-range contribution of the singular weights. 

Let us summarize our findings: Requirements for a successful treatment of nonlocal operators are twofold: 
\begin{itemize}
\item First, existing singularities must be taken into account correctly. Improper treatment results in a breakdown of the results for large $k$ in the high frequency domain.
\item Second, the long range interactions needs attention. Restricting the integration bounds results in an oscillatory behavior of calculated eigenvalues. Within this context
we may already mention: we will indeed observe oscillatory behavior
results based on the WKB-approximation compared to the high precision numerical results.
\end{itemize}

In the following we will demonstrate, that Heisenberg's  matrix representation of the nonlocal fractional Schr\"odinger equation is the ideal 
approach to fulfill both requirements adequately, which means both, an accurate treatment of the weakly singular kernel at $h=0$ as well as the consistent incorporation of its long range properties.

\section{The matrix representation of the stationary nonlocal Schr\"odinger equation in one dimension} 
In the one dimensional case using Dirac's bra-ket notation, the nonlocal stationary fractional Schr\"odinger equation is given according to (\ref{chxxsglfree}) by:
\begin{eqnarray}
\label{chxxsglfree00}
\tilde{\mathcal{H}}(w(|h|))  \psi &=&  
(\tilde{\mathcal{T}}(w(|h|)) + V)\psi = E \psi \\
&=&
(\tilde{\mathcal{G}}(w(|h|)) \otimes T + V)\psi = E \psi 
\end{eqnarray}
Introducing a complete set  of basis functions $\psi_n(x)$, the corresponding matrix representation is given by \cite{gre01b, messiah}:
\begin{eqnarray}
\label{chxxsglmat}
\braket{m|\tilde{\mathcal{H}}(w(|h|))|n} &=&  
\braket{m|\tilde{\mathcal{T}}(w(|h|)) + V|n} =\braket{m|E|n} \\
&=&
\braket{m|\tilde{\mathcal{G}}(w(|h|)) \otimes T + V|n} = \braket{m|E|n}\\
&=&
\label{chxxsglfree02}
\braket{m|\tilde{\mathcal{G}}(w(|h|) \otimes T|n}  + \braket{m|V|n} = \braket{m|E|n}
\end{eqnarray}
Eq. (\ref{chxxsglfree02}) immediately shows the major advantage of the matrix representation: kinetic energy and potential term may be treated separately. Once the nonlocal kinetic energy matrix elements are calculated, they may be used unmodified for many different potential types.

The nonlocal matrix elements for the kinetic energy are calculated in a three step procedure: 

\begin{enumerate}
\item apply the local kinetic energy operator, 
\item perform the nonlocalization integral,
\item realize the projection integral onto all elements on the set of basis functions $\ket{n}$. 
\end{enumerate}
This yields:
\begin{eqnarray}
\label{chxxsglmatTT}
\braket{m|\tilde{\mathcal{G}}(w(|h|) \otimes T|n} &=&
\int_{-\infty}^{\infty} \!\!\! dx \, \psi_m(x) 
\int_{-\infty}^{\infty}  \!\!\! dh \, w(|h|) s_h 
(-\frac{\hbar^2}{2 m}\partial_x^2) \psi_n(x)\nonumber \\
&&
\end{eqnarray}
where we introduced the shift operator $s_h$ with the property:
\begin{eqnarray}
\label{chxxshift}
s_h f(x)&=& f(x+h)
\end{eqnarray}
In the following, we use as a set of basis functions the harmonic oscillator basis, because of its superior asymptotic functional behavior. For reasons of simplicity, in the following we use natural units $\hbar = m = 1$.
  
 The harmonic oscillator basis $\psi_n(x)$  is generated by the solutions of the one dimensional Schr\"odinger equation of the harmonic oscillator in coordinate representation
\begin{equation}  
-\frac{1}{2} \partial^2_x \psi_n(x) = ( e_n -  \frac{1}{2} x^2  ) \psi_n(x) , \quad e_n = n+\frac{1}{2},  \quad n \in N_0 
\end{equation}  
and explicitly given by the set of normalized eigenfunctions 
\begin{eqnarray}  
\label{hob_psi}
\psi_n(x) = \frac{1}{\sqrt{\sqrt{\pi}2^n n!}} e^{-\frac{1}{2} x^2} H_n(x)\\
\langle m \mid n\rangle 
  = \int_\infty^\infty d x \psi_m(x)\psi_n(x)  = \delta_{n m }  
\end{eqnarray}  

Now we are able to calculate all matrix elements explicitly. In practical application it turns out, after performing the nonlocalization procedure that the remaining projection integrals are quite complicated to solve. For the special case of the Riesz weight, which we
will use to obtain the fractional kinetic energy term both integral operations commute. Therefore we will use a changed sequence of operations:

\begin{enumerate}
\item apply the local kinetic energy operator
\item perform the shift-operation only (not the nonlocalization integral) and then the projection integral  onto all elements on the set of basis functions $\ket{n}$ 
\item finally apply the nonlocalization integral with the Riesz-kernel. 
\end{enumerate}
This yields:
\begin{eqnarray}
\label{chxxsglmatTTcalc}
\braket{m|\tilde{\mathcal{G}}(w(|h|) \otimes T|n} &=&
\int_{-\infty}^{\infty}  \!\!\! dh \, w(|h|)  
\int_{-\infty}^{\infty} \!\!\! dx \, \psi_m(x) 
s_h (-\frac{\hbar^2}{2 m}\partial_x^2) \psi_n(x)\nonumber \\
&&
\end{eqnarray}
It should be noted that  this sequence of operations is much easier to handle:

It leads to well documented functions and as an unexpected bonus, the whole procedure may be  interpreted physically:  
It is now a weighted integral over shifted harmonic oscillator or Glauber-states \cite{cah69}. These states play a fundamental role
in solid states physics as a basic model for a quantum field theory of solids \cite{hak73}. In that sense, the Riesz-kernel forms
a specifically weighted infinite superposition of shifted harmonic oscillator states, which may be interpreted as an appropriate ansatz for a fractional  field theory of quantum liquids.
\index{Glauber states}
\index{shifted harmonic oscillator states}
\index{quantum liquid}

In the next section we will evaluate the integral (\ref{chxxsglmatTTcalc}) and derive a closed form for the Riesz weight fractional kinetic energy matrix elements.

\section{Matrix elements for the one-dimensional fractional kinetic energy term}
First we rewrite the classic kinetic energy term as a sum of eigenfunctions:

Using (22.7.13) and (22.8.7) from \cite{Ab} we obtain the recurrence relations for the eigenfunctions $\psi_n(x) \equiv \psi_n$
\begin{eqnarray}  
\label{hob_p2x}
x \psi_n                &=& \sqrt{\frac{n}{2}} \psi_{n-1} +  \sqrt{\frac{n+1}{2}} \psi_{n+1}\\
\label{hob_p2p}
\partial_{x} \psi_n &=& \sqrt{\frac{n}{2}} \psi_{n-1} -  \sqrt{\frac{n+1}{2}} \psi_{n+1}
\end{eqnarray}  
With (\ref{hob_p2x}) we obtain for the quadratic potential term:
\begin{eqnarray}  
\label{hob_x22}
x^2 \psi_n &=& 
\frac{1}{2} (
\sqrt{n (n-1)} \psi_{n-2} + (2 n+1) \psi_n +   \sqrt{(n+1)(n+2)} \psi_{n+2}) 
\nonumber\\
\end{eqnarray}  
With (\ref{hob_p2p}) we obtain for the second derivative:
\begin{eqnarray}  
\label{hob_p22}
\partial^2_{x} \psi_n &=& 
\frac{1}{2} (
\sqrt{n (n-1)} \psi_{n-2} - (2 n+1) \psi_n +   \sqrt{(n+1)(n+2)} \psi_{n+2}) 
\nonumber\\
\end{eqnarray}  
and the kinetic energy term results as:
\begin{eqnarray}  
\label{hob_kin0}
T  \psi_n(x)  &=& -\frac{1}{2} \partial^2_x\psi_n(x)\\
&=& -\frac{1}{4}  (
\sqrt{n (n-1)} \psi_{n-2} - (2 n + 1) \psi_n +   \sqrt{(n+1)(n+2)} \psi_{n+2})\nonumber\\
&&
\end{eqnarray}  
Applying the shift operation $s_h$ yields the shifted kinetic energy term 
\begin{eqnarray}  
\label{hob_kin1}
s_h T  \psi_n(x)  &=& -\frac{1}{4}  (
\sqrt{n (n-1)} \psi_{n-2}(x+h) - (2 n + 1) \psi_n(x+h) \nonumber \\
&& \qquad +   \sqrt{(n+1)(n+2)} \psi_{n+2}(x+h))
\end{eqnarray}  
The matrix element
\begin{eqnarray}  
\label{hob_kin111}
 \braket{m|s_h T|n}
  &=& -\frac{1}{4} \int_{-\infty}^{+\infty} dx \, \psi_m(x) \times\nonumber\\
&& (\sqrt{n (n-1)} \psi_{n-2}(x+h) - (2 n+1) \psi_n(x+h) \nonumber \\
&& \qquad +   \sqrt{(n+1)(n+2)} \psi_{n+2}(x+h))
\end{eqnarray} 
turns out to be equivalent to a sum of shifted harmonic oscillator or Glauber states $G_{m n}$ \cite{cah69}, which allows 
a new perception for an interpretation of  nonlocal operators, as we will demonstrate in the following. 
\index{Glauber states}
\index{shifted harmonic oscillator states}

With the abbreviation  
\begin{eqnarray}  
\label{hob_kin11}
G_{m \, n}(h) &=& \int_{-\infty}^{+\infty} dx \, \psi_m(x) \psi_{n}(x+h)\\
 &=& 
\sqrt{\frac{n!}{m!}}(-\frac{h}{\sqrt{2}})^{m-n}e^{-h^2/4}L_{n}^{(m-n)}(h^2/2)
\end{eqnarray} 
where $L_{n}^{(\beta)}(\xi)$ are the generalized Laguerre polynomials, we obtain for the matrix element for the shifted kinetic energy:
\begin{eqnarray}  
\label{hob_kin111}
 \braket{m|s_h T|n}
 &=&
  -\frac{1}{4} \bigl(  \sqrt{n (n-1)} \, G_{m\, n-2}(h) - (2 n+1) G_{m\, n}(h)\nonumber\\
  &&  \qquad+   \sqrt{(n+1)(n+2)} \, G_{m\, n+2}(h)\bigr) 
\end{eqnarray} 
In the next step we average over these states with a weight $w(h)$ to obtain the nonlocal kinetic energy matrix element:
\begin{eqnarray}  
\label{hob_kinnlx}
\braket{m|T_{nonlocal}|n}
 \int_{-\infty}^{\infty} \!\!\! dh w(h) \times  \braket{m|s_h T|n}
 \end{eqnarray} 
Different definitions of nonlocality therefore enter at this stage of evaluation via different choices of weights over 
coherent states.

We will use the Riesz definition $w(h) = |h|^{-\beta}$, which results in:
\begin{eqnarray}  
\label{hob_kinnl}
 \braket{m|T_{nonlocal}|n}
 &=& 
 \int_{-\infty}^{\infty} \!\!\! dh |h|^{-\beta}   \langle m \mid s_h T \mid n\rangle \\
 &=& 2 \int_{0}^{\infty} \!\!\! dh h^{-\beta}   \langle m \mid s_h T \mid n\rangle\\
\label{hob_kinnl3}
 &=& -\frac{1}{2} \sqrt{n (n-1)} \int_{0}^{\infty} \!\!\! dh h^{-\beta} G_{m\, n-2}(h)  \nonumber\\
 && \!\!  +\frac{1}{2}  (2 n+1)  \int_{0}^{\infty} \!\!\! dh h^{-\beta} G_{m\, n}(h)  \\
 && \!\! -\frac{1}{2}  \sqrt{(n+1)(n+2)} \int_{0}^{\infty} \!\!\! dh h^{-\beta} G_{m\, n+2}(h) \nonumber
\end{eqnarray} 
Let us evaluate these integrals using the explicit definition (13.6.9)  \cite{Ab} of the Laguerre polynomials $L_{n}^{(\tau)}(x)$:
\index{Laguerre polynomials}
\begin{eqnarray}  
\label{hob_kinnlv}
L_{n}^{(\tau)}(x)  &=& 
\frac{(1+\tau)_n}{n!} {_1}F_1(-n;1+\tau;x),  \qquad \tau > -1\\
  &=& 
\frac{\Gamma(1+\tau+n)}{\Gamma(1+\tau) \Gamma(1+n)}  
\sum_{j=0}^n    { \Gamma(-n+j) \over \Gamma(-n)}
                                { \Gamma(1+\tau) \over \Gamma(1+\tau+j)}{1 \over j!} x^j \nonumber \\
  &=& 
\frac{\Gamma(1+\tau+n)}{ \Gamma(1+n)}  
\sum_{j=0}^n    { \Gamma(-n+j) \over
\Gamma(-n) \Gamma(1+\tau+j)}{1 \over j!} x^j 
\end{eqnarray} 
and consequently
\begin{eqnarray}  
\label{hob_kinnl0}
\int_{0}^{\infty} \!\!\! dh h^{-\beta} G_{m\, n}(h) &=&
\sqrt{\frac{n!}{m!}} (-\frac{1}{2})^{\frac{m-n}{2}} \int_{0}^{\infty} \!\!\! dh e^{-h^2/4} h^{m-n-\beta} L_{n}^{(m-n)}(h^2/2) \nonumber \\
&=&
\sqrt{\frac{n!}{m!}} (-\frac{1}{2})^{\frac{m-n}{2}}\frac{\Gamma(1+m)}{ \Gamma(1+n)}  \times  \\
&&
\!\!\!\!\!\! \sum_{j=0}^n    { \Gamma(-n+j) \over \Gamma(-n)\Gamma(1+m-n+j)}{2^{-j} \over j!}   
\! \int_{0}^{\infty} \!\!\!\! dh \, e^{-\frac{h^2}{4}} h^{2 j + m-n-\beta}\nonumber
\end{eqnarray} 
Since the right hand integral is nothing else but a scaled $\Gamma$-function
\begin{eqnarray}  
\label{hob_kinn1l}
 \int_{0}^{\infty} \!\!\! dh  e^{-\frac{h^2}{4} } h^\sigma &=&
2^\sigma \Gamma(\frac{1}{2}(1+\sigma)) ,  \qquad \sigma > -1
\end{eqnarray} 
 we obtain 
\begin{eqnarray}  
\label{hob_kinnl2}
\int_{0}^{\infty} \!\!\! dh h^{-\beta} G_{m\, n}(h) 
&=&2^{-\beta}
\sqrt{\frac{m!}{n!}} (-\sqrt{2})^{m-n}    \frac{ \Gamma(\frac{1}{2}(1-\beta+m-n))}{\Gamma(1+m-n)} \times  \nonumber \\
&& 
\!\!{_2}F_1(\frac{1}{2}(1-\beta+m-n),-n;1+m-n;2)
\end{eqnarray}  
which may be inserted into (\ref{hob_kinnl3}) to obtain the final result: the explicit form for the kinetic energy matrix elements. 

We have derived the analytic result for the matrix-elements for kinetic energy term based on the Riesz definition of a fractional derivative in the
harmonic oscillator basis. It should be remarked, that the result contains only a finite number of terms in the series expansion of   the hyper geometric function ${_2}F_1(\frac{1}{2}(1-\beta+m-n),-n;1+m-n;2)$, since the second and third argument are negative integers. Consequently the matrix elements are known exactly and therefore the eigenvalues of the corresponding matrix may be calculated exactly in a finite number of steps. Only the limitations on the matrix dimension (finite $n,m$) cause an error in the resulting energy values. A concise error analysis will follow in the next section.

Moreover, it should be mentioned, that a corresponding derivation of a fractional kinetic energy matrix element based
on a different definition of a fractional derivative e.g. the Gr\"unwald-Letnikov derivative would lead to an infinite series of terms, since this is based on an infinite series expansion in terms of  standard derivatives of $n$-th order.

A first application of the derived results will be presented in the next section, where we calculate the energy spectrum
of the fractional Schr\"odinger equation with fractional harmonic oscillator potential. 

This is important, because 
the derived solution for the kinetic energy matrix elements may be simplified significantly, investigating the relation of the the fractional kinetic energy term and its Fourier-transform.   

\section{The one-dimensional fractional oscillator}
The harmonic oscillator potential $V(x) = \frac{1}{2} \omega x^2$ plays an outstanding role in quantum physics. 

First, it serves
as the starting point for a discussion of vibrational problems, since it generates the ideal equidistant level spacing for a vibrational spectrum, used in all branches of physics. 

Second, it serves as a valid approximation for any kind of potentials, where we are interested in the low
excitation region near a global or local minimum. 

Last not least, it plays an outstanding role for an understanding of the
particle-wave duality in the quantum world, since it is connected  to the kinetic energy operator by a simple Fourier transform.
A quantum state may be described in coordinate space, where the coordinate representation of the operators for position $\hat{x}$ and momentum $\hat{p}$ are given by 
\begin{eqnarray}  
\label{hob_opsc}
\hat{x}&=& x\\
\hat{p}&=& i \hbar \partial_x
\end{eqnarray} 
and the Schr\"odinger equation in coordinate representation
\begin{eqnarray}  
\label{hob_qsx}
H \psi(x) = ( -\frac{1}{2}\frac{\hbar^2}{m^2}\partial_x^2 + V(x) ) \psi(x) = E \psi(x)
\end{eqnarray} 
as well as in momentum space, where the momentum representation of the operators for position $\hat{x}$ and momentum $\hat{p}$ are given by  
\begin{eqnarray}  
\label{hob_opsc}
\hat{x}&=& i \hbar \partial_k\\
\hat{p}&=& k
\end{eqnarray} 
and the Schr\"odinger equation in momentum representation
\begin{eqnarray}  
\label{hob_qsk}
H \psi(k) = ( -\frac{1}{2}\frac{\hbar^2}{m^2}\partial_k^2 + V(k) ) \psi(k) = E \psi(k)
\end{eqnarray} 
Both approaches yield the same results in terms of the same energy level spectrum $E$. 

The Fourier-transform of the derivative operator is just $k$.  From all possible oscillator potentials of type $V(x) = \frac{1}{2} \omega  (x^2)^{\sigma/2}$,  only for the harmonic oscillator ($\sigma=2$)  we obtain a complete symmetry in $\{x,k\}$  (for reasons of readability we set $\hbar=m=\omega=1$ from now on):
\begin{eqnarray}  
\label{hob_qsx}
H \psi(x) &=& (-\frac{1}{2}\partial_x^2 + \frac{1}{2} x^2 ) \psi(x) = E \psi(x)\\
H \psi(k) &=& (-\frac{1}{2}\partial_k^2 + \frac{1}{2} k^2 ) \psi(k) = E \psi(k)
\end{eqnarray} 
Does this  this symmetry still hold for a nonlocal extension of quantum mechanical operators? In addition the question arises, which symmetries are present in a nonlocal extension of the standard Schr\"odinger equation at all?

\begin{figure}[t]
\begin{center}
\includegraphics[width=14cm]{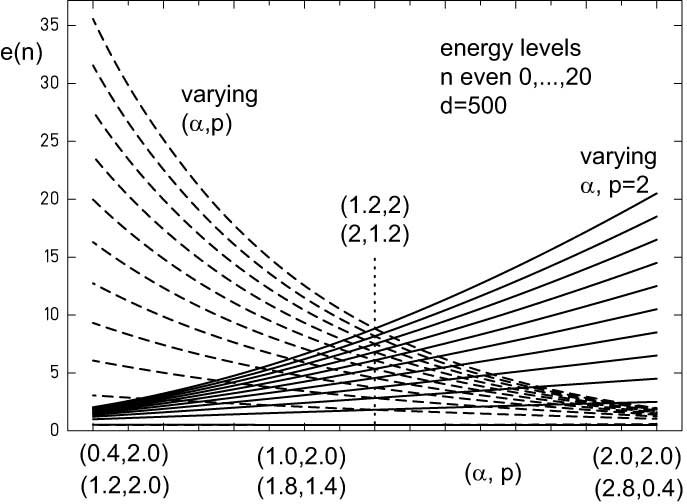}    
\caption{Energy levels for $n=0,...,20$. Thick lines indicate running $\{\alpha, p=2\}$ (abzissa: first line). 
Dashed line indicates running $\{\alpha, p\}$ (abzissa:second line). At $\{\alpha = 1.2, p=2\}$ 
and its Fourier transform $\{\alpha = 2, p=1.2\}$ (pointed line) both spectra coincide exactly. 
}
\label{fig9}
\end{center}
\end{figure}

We will discuss this question for the  Riesz-weight, because for the special case of the of the Laplacian in one dimension it is equivalent to the fractional derivative of formal type $\partial_x^{\alpha/2} \equiv  (-\Delta)^{\alpha/4}$ with $0 \leq \alpha \leq 2$.
We formally may set in coordinate space
\begin{eqnarray}  
\label{hob_opsrrr}
\hat{x}&=& |x|^{\alpha/2}\\
\hat{p}&=& \partial_x^{\alpha/2}
\end{eqnarray} 
and formally in momentum space
\begin{eqnarray}  
\label{hob_opsrrm}
\hat{x}&=& \partial_k^{\alpha/2}\\
\hat{p}&=& |k|^{\alpha/2}
\end{eqnarray} 
As a consequence, once again we obtain two representations for a fractional Schr\"odinger equation in coordinate and momentum space.  

Let us consider 
right from the beginning a broader range of possible oscillator  potentials:  
\begin{eqnarray}  
\label{hob_qsrxrr}
H \psi(x) &=& ((-\Delta)^{\alpha/2} +  (x^2)^{p/2}) \psi(x) = 2 E \psi(x),   \\
             &&  \qquad\qquad 0 \leq \alpha \leq 2,  \, p\geq 0,  \, \alpha,p \in \textrm{R}\nonumber \\
H \psi(k) &=& ((-\Delta)^{p/2} + (k^2)^{\alpha/2}) \psi(k) = 2 E \psi(k),  \\
             &&  \qquad\qquad 0 \leq p \leq 2, \, \alpha\geq 0, \, \alpha,p \in \textrm{R}  \nonumber
\end{eqnarray} 
or in matrix representation 
\begin{eqnarray}  
\label{hob_qsrxrr2}
\braket{m|H|n} &=& \braket{m|(-\Delta)^{\alpha/2} +  (x^2)^{p/2}|n} = 2 \braket{m|E|n}, \\
\braket{m|H|n} &=& \braket{m|(-\Delta)^{p/2} +  (k^2)^{\alpha/2}|n} = 2 \braket{m|E|n}, \\
                       &&  \qquad\qquad \alpha\geq 0,\, p\geq 0, \, \alpha,p \in \textrm{R}\nonumber
\end{eqnarray} 
which for $\alpha=2$ and $p=2$ reduces to the standard quantum harmonic oscillator.
The matrix elements 
 $\braket{m|(-\Delta)^{\alpha/2}|n}$ 
 in the harmonic oscillator basis
 are identical with the fractional kinetic energy matrix elements 
$ \braket{m|T_{nonlocal}|n}$ in (\ref{hob_kinnl}) and have been calculated in the previous section. 

\begin{figure}[t]
\begin{center}
\includegraphics[width=\textwidth]{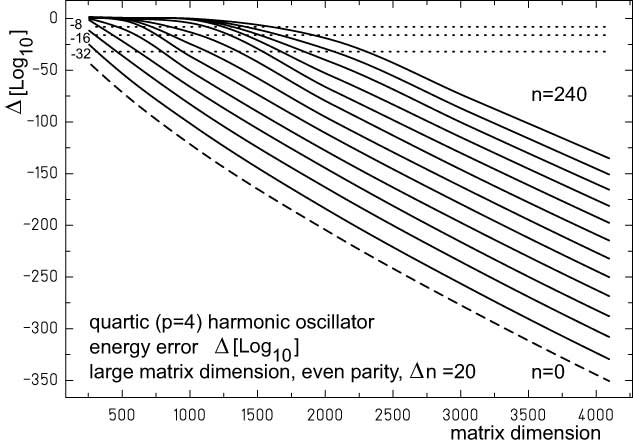}    
\caption{For the quartic ($p = 4$) harmonic oscillator energy levels $e_n$  the absolute error $\Delta[\log_{10}]$ depending on the matrix dimension is shown for level number $n$ in the range  from $n=0$ (ground state) to $n=240$ in $20$ steps. Dotted lines in
the upper part mark accuracy thresholds $\Delta=10^{-16}$ and $\Delta=10^{-32}$ respectively, which correspond to accuracy goals in numerical calculations limited on  standard double, long double or quadruple number representation given by specific compiler implementations.    
}
\label{fig3}
\end{center}
\end{figure}

It will turn out, that both equations will generate the  same energy spectra for a given set $\{ \alpha, p\}$. 
In addition, due to the
use of the harmonic oscillator basis and its asymptotic behavior, the calculation of the matrix elements for fixed values of  $\alpha$ and $p$ converges for any $\alpha >0$ and $p>0$.

Hence for the special case of the Riesz-weight with its special Fourier transform, the
concept of wave-particle duality still holds for the fractional case.
\index{duality}

Even more exciting, this duality leads to a generalized symmetry, which manifests itself very prominently using the matrix representation of the fractional 
Schr\"odinger equation, as we will demonstrate in the following.

The arguments listed above emphasize the fundamental importance to obtain matrix elements for the fractional  oscillator potential
\begin{eqnarray}  
\label{hob_matVVV}
V(x) = |x|^p,  \qquad  \qquad p \in \textrm{R}
\end{eqnarray} 
where the absolute value guarantees positive parity and the power $p$ as a positive real number.
\index{fractional harmonic oscillator potential}

Potentials of this type have been considered widely in literature. Especially in quantum chemistry or nuclear physics potentials of this type 
have been used to model e.g. vibrational behavior of molecules or properties of phonons in solid state physics \cite{bel45, sif49, messiah}. The corresponding Schr\"odinger equation 
was mostly solved using approximate methods like  WKB-approximation or the Ritz variational principle and with the restriction to integer p values $p \in \textrm{N}$.
 
In a series of seminal papers Palma and coworkers \cite{pal87} derived a closed formula for the potential energy matrix elements restricted to integer values of $p$ based on the combined use of Cauchy's 
integral formula and the Baker-Campbell-Hausdorff theorem for harmonic oscillator matrix elements.
Using the notation $p = 2 \sigma$ and  with the abbreviation $\mu = \sigma + (m - n)/2$ they deduce
\begin{eqnarray}  
\label{hob_potint}
 \braket{m|x^{2 \sigma}|n}  &=&
\sum_{i=0}^{\max(m,\mu)} \frac{\sqrt{m! n!} (2 \sigma)!}
{i! (m - i)! (n - m + i)! (\mu-i)! \,2^{ \sigma + \mu - i} }\nonumber \\
& & \qquad \quad\,\,\, \sigma \in \textrm{N}, \quad m-n = \textrm{even}, \quad m\geq n
\end{eqnarray} 
which of course is an ideal starting point to extend this form from integer $p$ to real $p$. We obtain
\begin{eqnarray}  
\label{hob_potreal}
 \braket{m| |x|^{2 \sigma}|n}  &=&
\sum_{i=0}^{\max(m,\lceil\mu\rceil)} \frac{\sqrt{m! n!}\,  \Gamma(1+2 \sigma)}
{i! (m - i)! (n - m + i)! \,   \Gamma(1+\mu-i)  \,2^{ \sigma + \mu - i}}\nonumber \\
& & \qquad  \qquad\sigma \in \textrm{R}, \quad m-n = \textrm{even}, \quad m\geq n
\end{eqnarray} 
where $\lceil  \rceil$ denotes the  ceiling-function. 
\index{ceiling-function}
\index{functions!ceiling}
\index{functions!Gauss-bracket}

Once the matrix elements for the kinetic and potential energy term are known, we may calculate the eigenfunctions and eigenvalues of 
the fractional Schr\"odinger 
equation (\ref{hob_qsrxrr2}) with the fractional oscillator potential (\ref{hob_matVVV}) by diagonalization of its matrix representation.

Numerical results for the ground state eigenvalues are shown in figure \ref{fig8} for increasing matrix dimension. 

The
relative error  for ground state energy $\Delta e_0$ and first excited state $\Delta e_1$ with respect to the solution for $d=512$ is plotted in figure \ref{fig6a} and figure \ref{fig6b}. 

\afterpage{
\begin{figure}[p]
\begin{center}
\includegraphics[width=11cm]{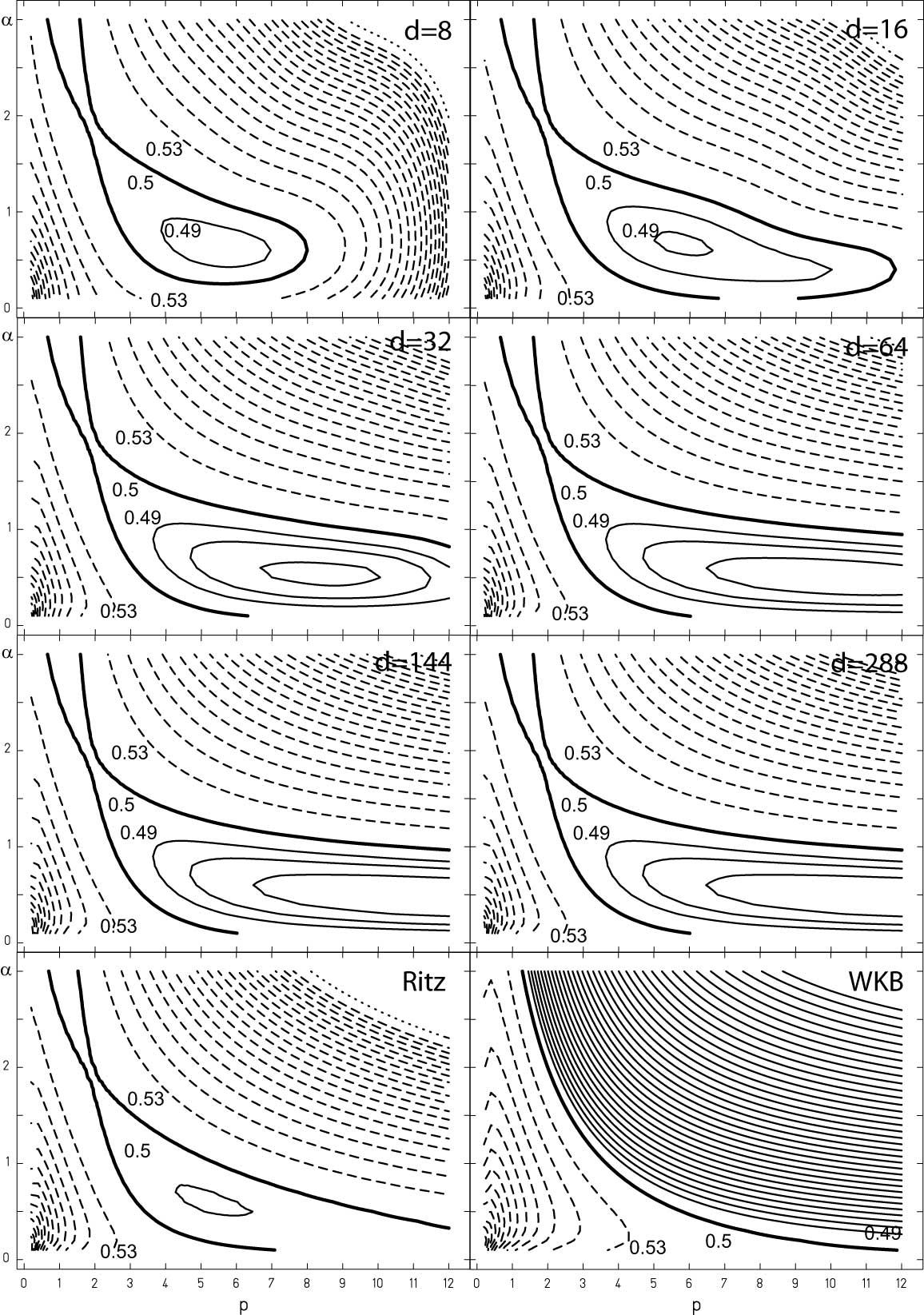}    
\caption{Contour plot of the zero point energy (lowest eigenvalue $n=0$) of the fractional Schr\"odinger equation with harmonic oscillator potential for increasing values of matrix dimension $d$. Solid thick line indicates $e_0 = 0.5$ Solid thin lines indicate $e_0 < 0.5$ in $-0.01$ steps. Dashed lines indicate $e_0 > 0.5$ in $+0.03$ steps.  In addition, Ritz- and WKB-approximate solutions are plotted.  
}
\label{fig8}
\end{center}
\end{figure}
\clearpage
}

\afterpage{
\begin{figure}[p]
\begin{center}
\includegraphics[width=11cm]{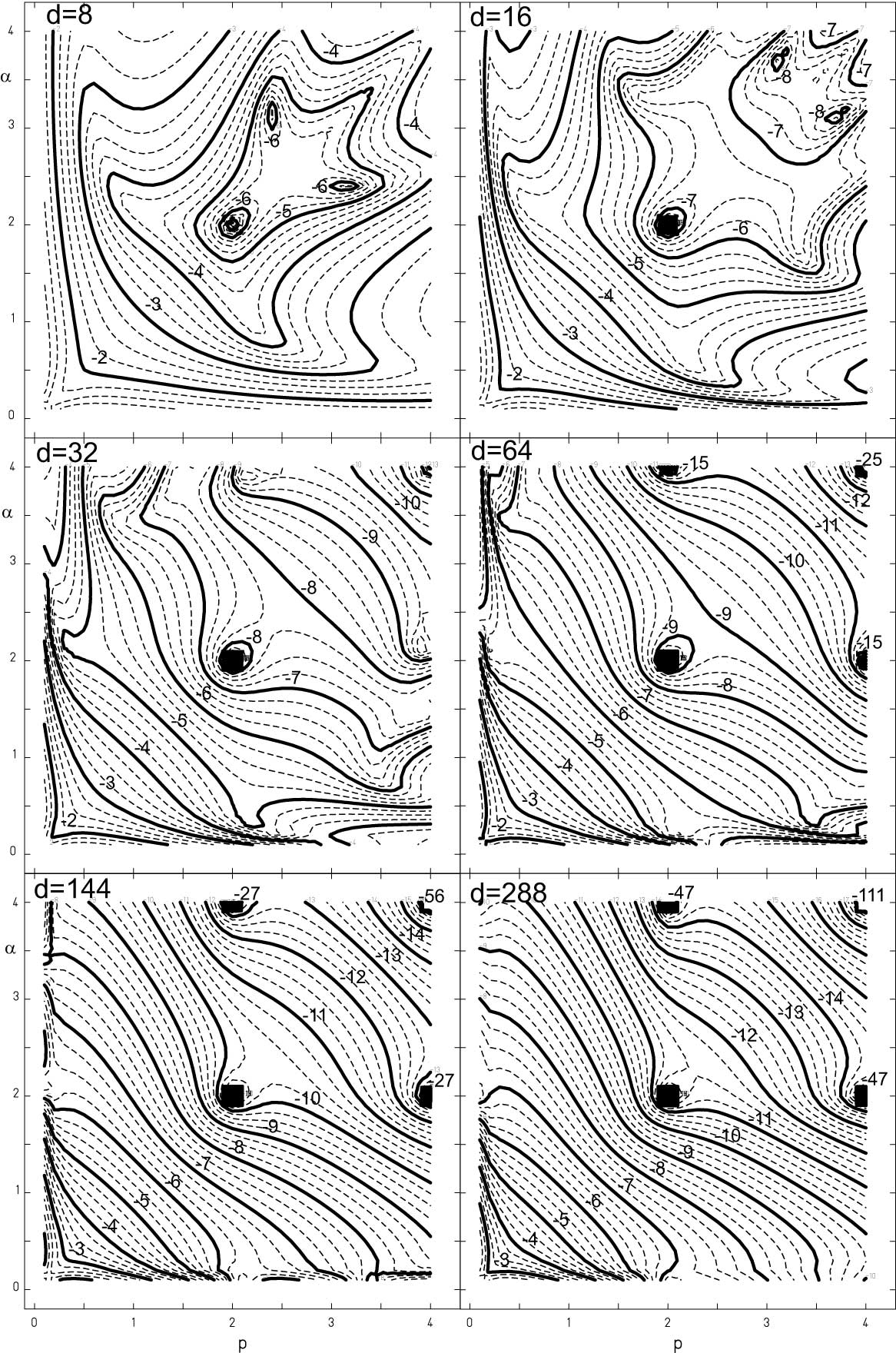}    
\caption{Contour plot of the $\log_{10}$ relative error of the zero point energy ($\Delta e_0$) of the fractional Schr\"odinger equation with harmonic oscillator potential for increasing values of matrix dimension $d$.  Dashed lines increment $\delta = 0.25$. Squares indicate local minima  (light houses) for the error.
}
\label{fig6a}
\end{center}
\end{figure}
\clearpage
}

\afterpage{
\begin{figure}[p]
\begin{center}
\includegraphics[width=11cm]{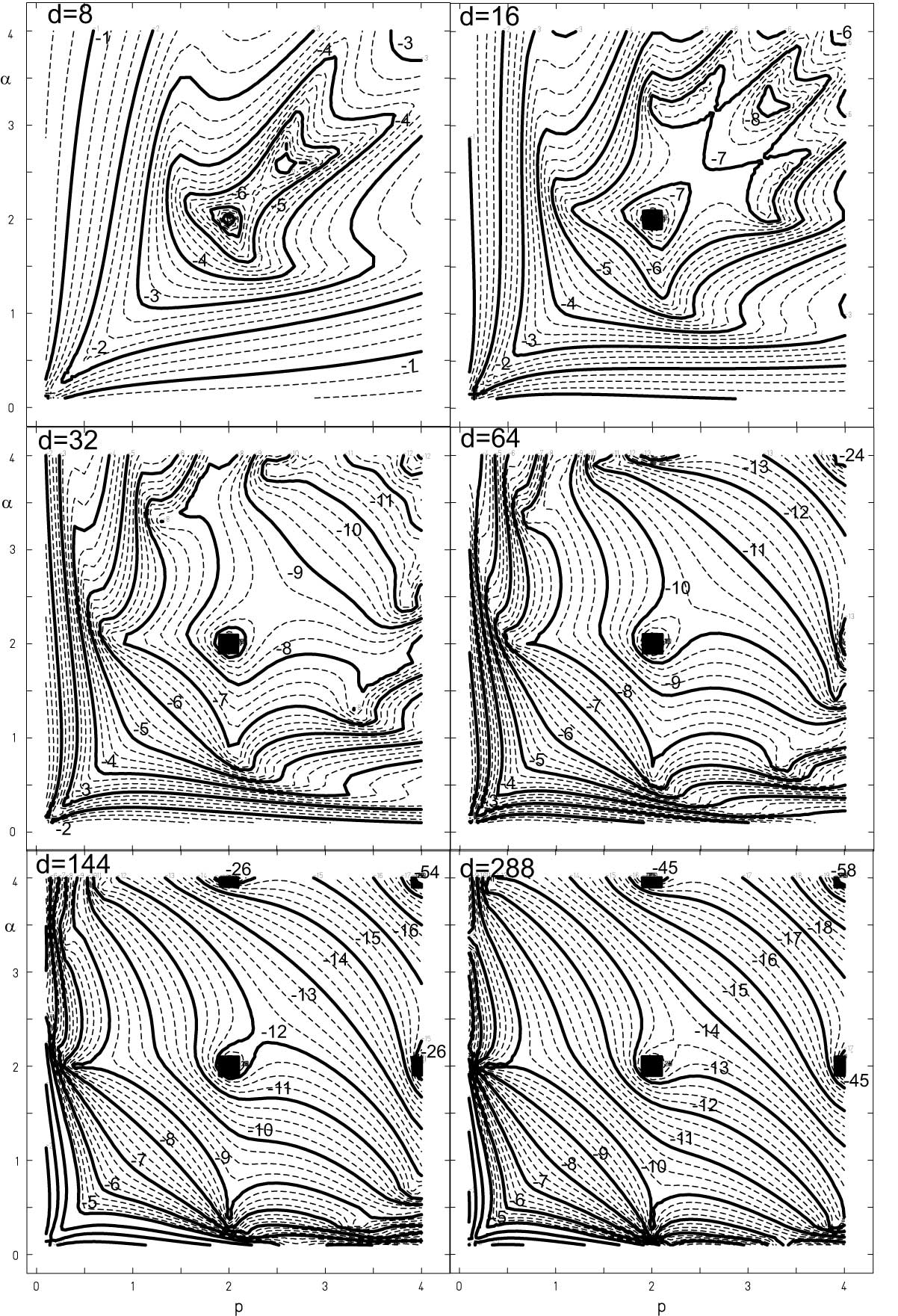}    
\caption{Contour plot of the $\log_{10}$ relative error of the energy of the first excited state ($\Delta e_1$) of the fractional Schr\"odinger equation with harmonic oscillator potential for increasing values of matrix dimension $d$. Dashed lines increment $\delta = 0.25$.   Squares indicate local minima (light houses) for the error. 
}
\label{fig6b}
\end{center}
\end{figure}
\clearpage
}
Increasing the matrix dimension $d$ should yield 
increasing accuracy of the results. Indeed we observe a rapid increase of accuracy in the region near $\{\alpha=2, p=2\}$. 
A stabilization of energy values on the plane $\{0 < \alpha <3, 0<p<12\}$, which means, a relative  error less than $0.001$ is reached for values of $\alpha$
and $p$ in the vicinity of $\{\alpha=2, p=2\}$ for matrix dimensions $d>16$. Only for values $\alpha \ll1$ and $p \ll 1 $ the matrix dimension must be chosen $d>144$ in
order to acquire a similar accuracy. 

Remarkable in figure \ref{fig6a} and figure \ref{fig6b} is the emergence of light houses in the error plane, characterized by extremely high precision results compared to the error in the immediate neighborhood
region, which become pronounced for increasing matrix dimension.  These points are characterized by even integer positions $\{ \alpha = 2n, p = 2m  \}$ $n,m \in N_0$.

E.g. from figure \ref{fig6a} we obtain for the ground-state energy error at $\{\alpha=4,p = 4\}$ a value of   $\Delta e_0 = 10^{-111}$ using a matrix dimension $d=244$,
while already in the close neighborhood the same error is not better than $10^{-20}$.
\index{light houses}
A reasonable explanation results from the number of off-diagonal elements for a given position in the $\{\alpha,p\}$ plane:

At these distinct positions, the matrix becomes sparse with a small and finite number of off-diagonal matrix elements, while all other position pairs yield the full set of all possible
off diagonal elements. 
  
In contrast to the classical harmonic oscillator this fractional Schr\"odinger
equation has not been solved analytically until now. Nevertheless there are different methods to derive approximate analytic expression for the energy levels, which we will use in the following to check the reliability of our results and to 
investigate the limitations of such approximate solutions. 

In the next section we will therefore present approximate solutions based on  the WKB-approximation and Ritz-variational principle and we will 
compare them with the numerical results.  

\section{Analytical approximations: WKB and Ritz variational principle }
Based on the  Bohr-Sommerfeld quantization rule Laskin has derived  an analytic expression \cite{laskin} for the
energy values of the fractional quantum harmonic oscillator, which applies the WKB-approximation to the fractional case.

It is given by
\begin{equation}
\label{ch14eho122}
E_{WKB}(n,\alpha, p)  =  
\frac{1}{2}
\left(
(n +\frac{1}{2})
{\pi p \over 2 B (\frac{1}{p}, \frac{1}{\alpha} +1)}
\right)^\frac{\alpha p}{\alpha+p}  
\quad n=0,1,2,...
\end{equation}
where $B(x,y)$ is the beta function or Euler integral of the first kind.
\index{functions!Beta-function}

Within the WKB-formalism the wave-function is approximated by a combination of separately determined functions. First they are derived within the classically allowed and forbidden regions  independently from each other. Then they are stitched together at the classical turning points to obtain
a global approximate solution. This is of course in contradiction to the global ansatz for a nonlocal approach like in fractional calculus, where
the long-term interaction may not be cut off arbitrarily at a classical turning point.  

A comparison of energy levels with the numerical solutions is presented in figure \ref{fig7}, which shows the relative error for the four lowest eigenvalues. 
For the ground state energy $e_0$ this error easily reaches $20\%$, which is very bad. For higher excitations the error behaves increasingly better (e.g. for $e_3$ the error
reduces to $0.3\%$).
Depending on the region of
$\{\alpha,p\}$ positions, the WKB-approximate energies over- or underestimate the high precision numerical results, which is a behavior, which we discussed in the
last section, characteristic for a finite cutoff in the region of integration of the non-local extension of local operators. 
\index{WKB-approximation}
Thus we obtain results, which are reliable only for higher energy levels and which oscillate around the exact value.  

Requirements for a better approach are twofold: accuracy should be increased for the lowest energy levels and the approximation should reliably overestimate the exact
result, giving a reliable upper bound for the exact solution.

These demands are fulfilled by an
alternative approach to obtain approximate solutions for the fractional harmonic oscillator energy levels and functions based on the  Ritz' variational method,
which we will introduce as an independent, alternative way to obtain analytical approximations, which may be compared with the numerical results:

\index{Ritz variational method}

\begin{figure}[t]
\begin{center}
\includegraphics[width=\textwidth]{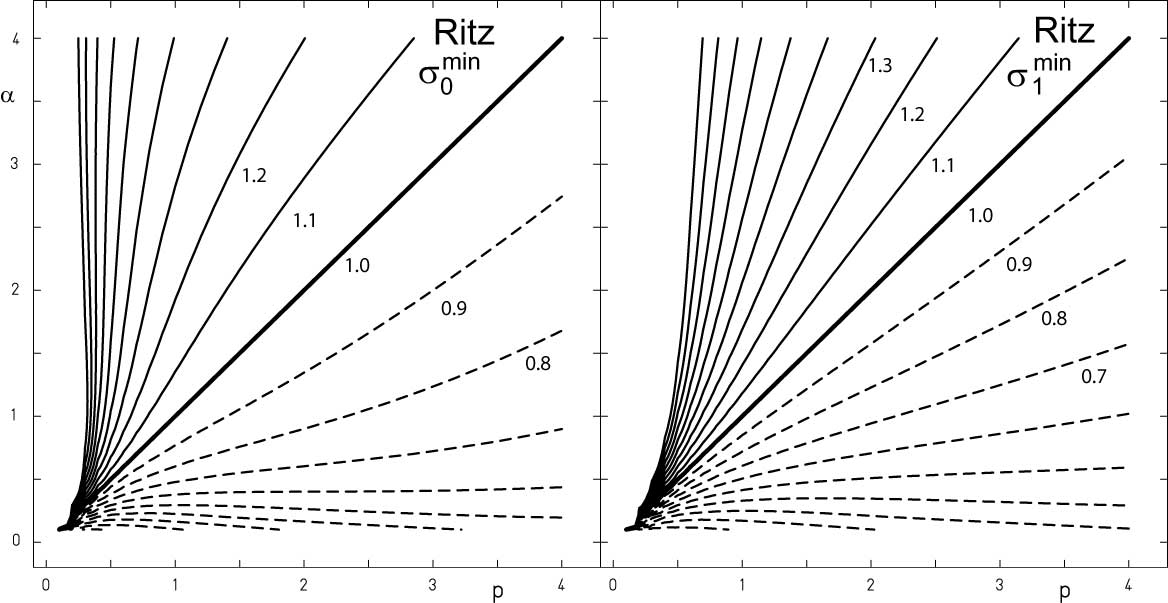}    
\caption{Optimum inverse width $\sigma^{min}_i$ or variance root for the Ritz variational Gaussian test functions $ \Psi_0^+(\sigma) $ (left side) and $\Psi_1^-(\sigma)$ (right side), minimizing the energy functional (\ref{hob_testfun1}).
}
\label{figsigma}
\end{center}
\end{figure}
\begin{figure}[t]
\begin{center}
\includegraphics[width=\textwidth]{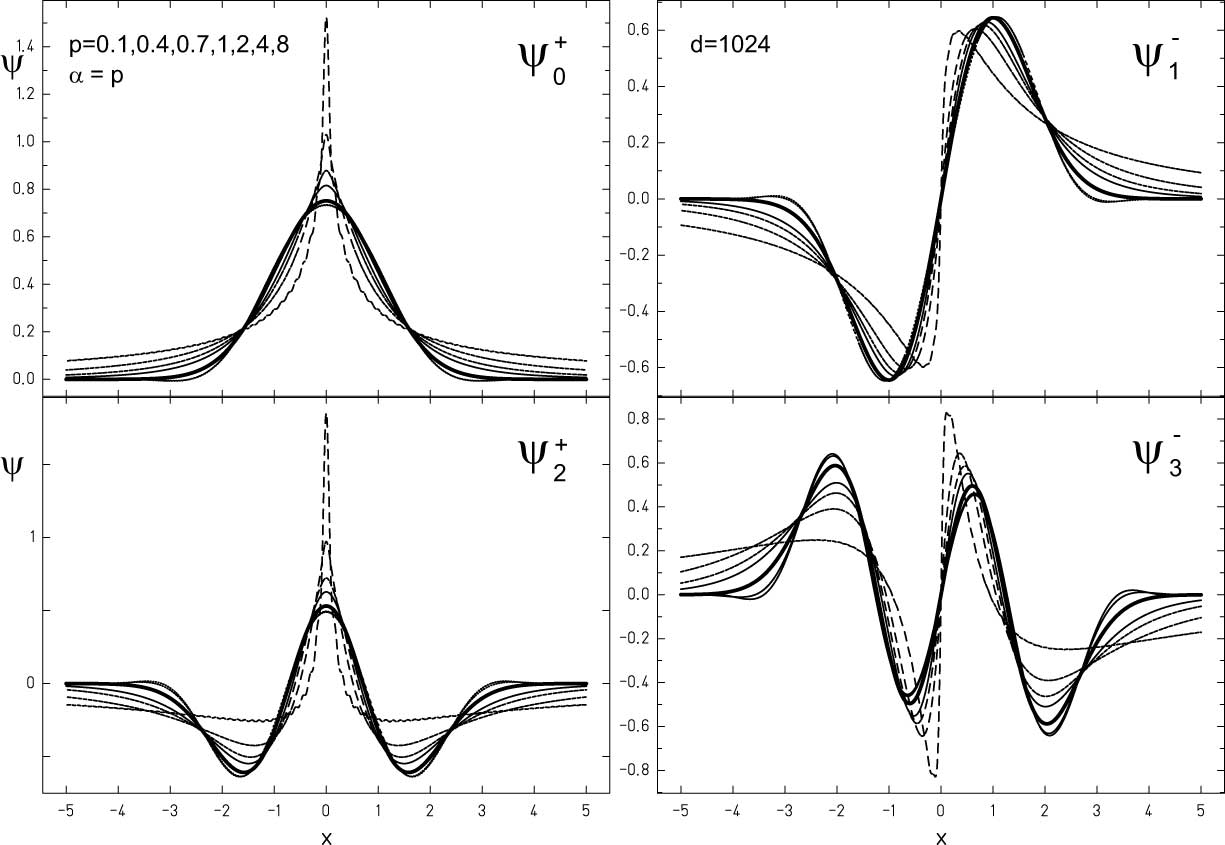}    
\caption{The four lowest eigenfunctions, of the fractional harmonic oscillator along the $\sigma=1$ - line, which corresponds to the case $\alpha=p$. Thick line
indicates the classical $\alpha=p=2$, dashed lines indicate $\alpha=p < 2$ and show a general tendency for a increasing concentration at the origin  for smaller $\alpha=p$ values. Pointed lines indicate the graphs for $\alpha=p > 2$.   
}
\label{figwellpl1}
\end{center}
\end{figure}

\begin{figure}
\begin{center}
\includegraphics[width=10.5cm]{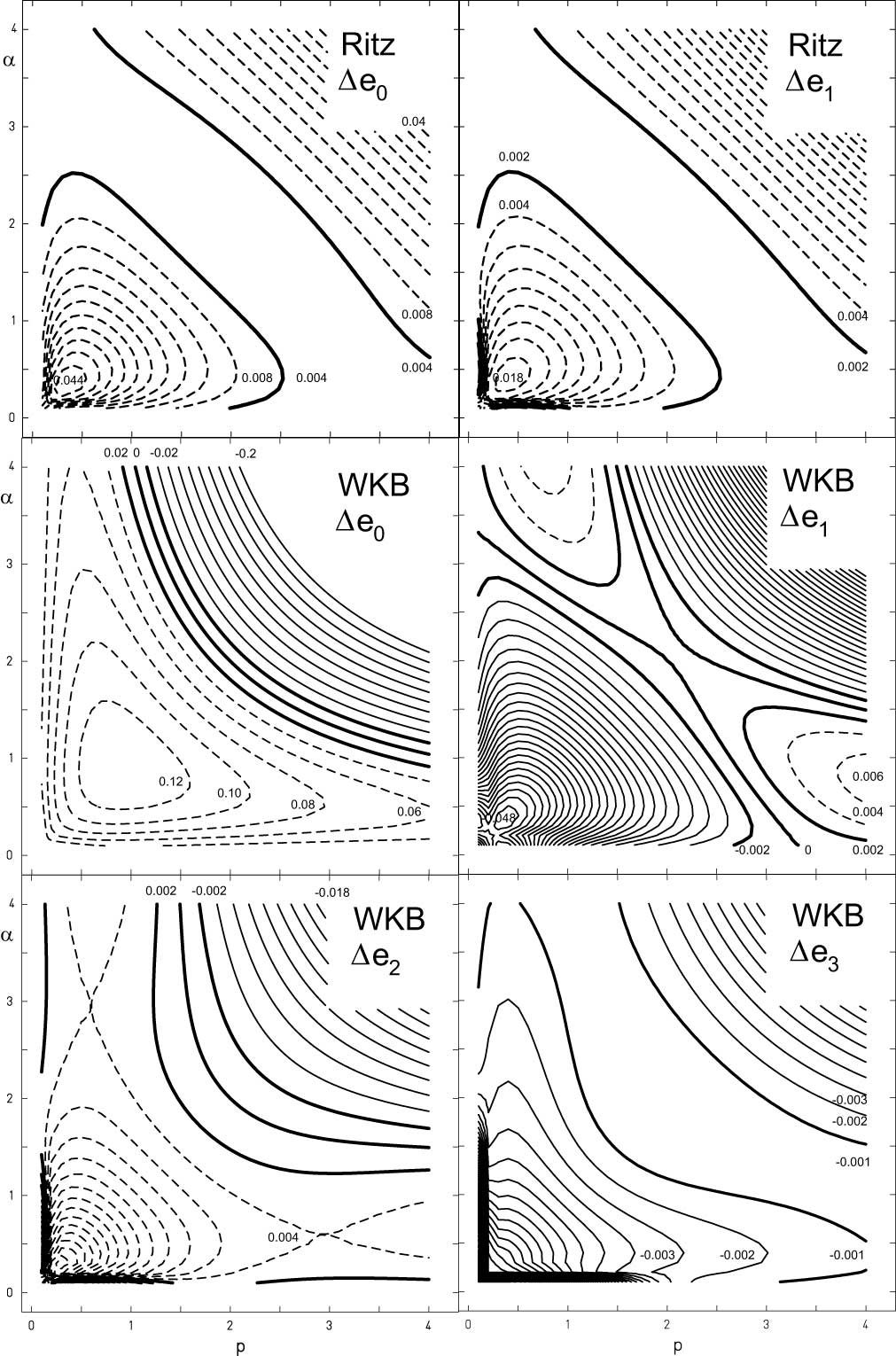}    
\caption{Contour plot of the relative  error of the energy of ground state (eigenvalue $n=0$) and the first excited state (eigenvalue $n=1$) of the fractional Schr\"odinger equation with harmonic oscillator potential for Ritz variational result and WKB approximation respectively compared to the exact diagonalized eigenvalues (matrix dimension $d=288$). Thin dashed lines show overestimation, thin solid lines show underestimation of the numerical solution.
}
\label{fig7}
\end{center}
\end{figure}

We will use the following two test functions, which extend the even parity ground state and the odd parity first 
excited state of the standard harmonic oscillator eigenfunctions introducing a variable inverse width $\sigma$ or 
variance $\sigma^2$:   
\begin{eqnarray}  
\label{hob_testfun}
 \Psi_0^+(\sigma) &=& \frac{1}{\sqrt{\pi}} e^{-\frac{1}{2}{x^2}/{\sigma^2}}\\
 \Psi_1^-(\sigma) &=& \sqrt{\frac{2}{\sigma^3 \sqrt{\pi}}} \, x e^{-\frac{1}{2}{x^2}/{\sigma^2}}
 \end{eqnarray} 
which will be adjusted such, that the energy expectation value of the energy functional
\begin{eqnarray}  
\label{hob_testfun1}
 E(\sigma) &=& \frac{ \braket{\Psi|H|\Psi}}{ \braket{\Psi|\Psi}}
 \end{eqnarray} 
is minimized.
From the condition
\begin{eqnarray}  
\label{hob_testfun2}
 \frac{dE(\sigma)}{d \sigma} &=& 0
 \end{eqnarray} 
 we may deduce the corresponding variance roots minimizing the ground state energy:
 \begin{eqnarray}  
\label{hob_testfun4}
 \sigma_0^{\min} 
  &=& (\frac { \alpha\Gamma(\frac{1}{2}+\frac{\alpha}{2})}{p\,  \Gamma(\frac{1}{2}+ \frac{p}{2})})^{p+\alpha}
 \end{eqnarray} 
 and minimizing the first excited state energy:
 \begin{eqnarray}  
\label{hob_testfun6}
 \sigma_1^{\min} 
 &=& (\frac { \alpha\Gamma(\frac{3}{2}+ \frac{\alpha}{2})}{p\,  \Gamma(\frac{3}{2}+ \frac{p}{2})})^{p+\alpha}
 \end{eqnarray} 
The functional behavior of the optimum width is sketched in figure~\ref{figsigma}. 

While increasing $\alpha$ results in a broadening of the test function the increase of the potential power $p$ yields a narrowing. 

For the special case 
$\alpha=p$ the optimum
wave function is characterized by a direct compensation of these two effects, leaving the shape of the test functions unchanged compared to the standard quantum harmonic oscillator eigenfunctions ($\alpha=p=2$). 

A comparison
with the high precision numerical solutions is given in figure  \ref{figwellpl1}. 

We recognize, that the Ritz test functions are a good approximation in the case $\alpha=p$, 
especially in the region $\alpha=p > 2$. Significant deviations occur for $\alpha=p \ll 2$, where numerical solution tend to be increasingly concentrated around the origin.
 
With these settings, we obtain for  the ground state energy $E_0$ 
\begin{eqnarray}  
\label{hob_testfun3}
 E_0(\sigma) &=&  \frac{\braket{ \Psi_0^+|T+V| \Psi_0^+}}{\braket{ \Psi_0^+| \Psi_0^+}} \\
  &=&  \frac{\Gamma(\frac{1}{2}+\frac{\alpha}{2})}{2\sqrt{\pi}\, \sigma^{\alpha}} +
  \frac{\Gamma(\frac{1}{2}+\frac{p}{2})\, \sigma^p}{2\sqrt{\pi}}
 \end{eqnarray} 
 and the first excited odd state:
 \begin{eqnarray}  
\label{hob_testfun5}
 E_1(\sigma) &=&  \frac{\braket{ \Psi_1^-|T+V| \Psi_1^-}}{\braket{ \Psi_1^-| \Psi_1^-}} \\
  &=&  \frac{\Gamma(\frac{3}{2}+\frac{\alpha}{2})}{\sqrt{\pi}\, \sigma^{\alpha}} +
  \frac{\Gamma(\frac{3}{2}+\frac{p}{2})\, \sigma^p}{\sqrt{\pi}}
 \end{eqnarray} 
A comparison with the numerical solutions is presented in figure \ref{fig7}, which shows the relative error for the four the ground state ($n=0$) and first excited state ($n=1$). The overall agreement is very good with a maximum error of about $4\%$. Since the Riesz test function covers only a part of possible variations, the energy is always overestimated, a feature, already known for local problems.

Comparing the WKB-approximation and the Ritz-method with the numerical solutions, we may deduce the following results:
For the lowest eigenvalues $e_0 , e_1$ the WKB-approximations yields very unprecise results, while the Ritz variation leads to much better results, even with the simple test-function, we have used. For higher energy values $e_n$ the WKB-error decreases. Another feature is the oscillatory sign-change behavior of the WKB-error, which we
explain with the inconsistency of using piece wise continuous functions in a globally  nonlocal scenario.

\begin{figure}[t]
\begin{center}
\includegraphics[width=\textwidth]{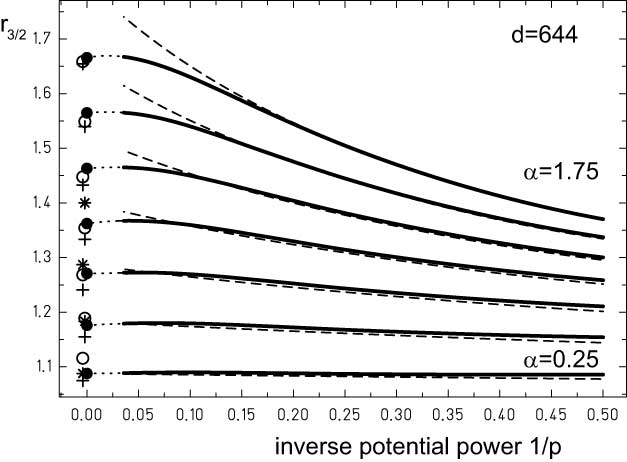}    
\caption{Ratio of eigenvalues $r_{3/2}=e_3/e_2$ of the fractional oscillator for different $\alpha$  in the range $0.25 \leq \alpha \leq 1.75$ in 0.25 steps and increasing power $p$ of potential $|x|^p$ plotted as a function of $1/p$ . The limit $p \rightarrow \infty$ is the infinite potential well. First, diagonalization was done for a sequence of powers $2 \leq p \leq 10$ (numerical solution given with  thick lines), then ratios were extrapolated for the limit $1/p \rightarrow 0$ ($r^{\textrm{fit}}_{3/2}$ dotted lines and thick bullet at $1/p=0$). Dashed lines indicate expected energy ratios $r^{\textrm{WKB}}_{3/2} $ calculated with WKB-approximation (\ref{hob_rWKB}), which turn increasingly worse for larger $p$. Especially near $\alpha=2$ which corresponds to the local limit, the deviations become larger for $p \rightarrow \infty$ and we detect the weakness  of the WKB-solutions, caused by the wrong asymptotic behavior. \newline
At $1/p=0$ (the potential well limit) the ratio for WKB-extrapolation $r^{\textrm{WKB}}_{3/2}(p \rightarrow \infty) $ is given by a star symbol,  hollow circles show the ratio $r^{\textrm{class}}_{3/2}$ according to the simple $k$ power law (\ref{hob_rclass}) and crosses indicate calculated approximations $r^{\textrm{cs}}_{3/2}$ following (\ref{hob_rcs}), bullets indicate the extrapolation of the numerical result. Near  the local limit $\alpha=2$ all different strategies lead to the same results, differences occur 
for $\alpha \ll 2$.   
}
\label{figwall}
\end{center}
\end{figure}

\section{A limiting case: The infinite potential well}
\index{infinite potential well}
The infinite potential well serves as an ideal candidate to apply basic concepts of fractional calculus, where we are able to investigate the influence of nonlocal concepts in general. 
Centered at the origin with finite size $2$, it is explicitly given by: 
\begin{numcases}
{V_{\textrm{wall}}(x) =} 
0         &  $|x| \leq 1$\\
   \infty &   $|x|>1$
\end{numcases}
Since this potential may be considered as the limiting case of the harmonic oscillator potential with increasing potential power $p$:
\begin{equation}  
\label{hob_wallLimit}
V_{\textrm{wall}}(x) = \lim_{p \rightarrow \infty} \frac{1}{2} |x|^p ,
 \end{equation} 
the spectrum of the fractional Schr\"odinger equation with infinite potential well may serve as a guide to check the validity of our approach  to solve the
 fractional Schr\"odinger equation ( \ref{hob_qsrxrr2} ) for the
fractional harmonic oscillator potential in the limit $ \lim_{p \rightarrow \infty}$.
Furthermore we may compare obtained numerical results with several analytic approximations fore the level spectrum given in literature.

We first calculated the eigenvalue spectrum $e_n $ of the fractional harmonic oscillator numerically  for increasing potential power $p$, $2 \leq p \leq 10, \Delta p = 2$  and defined the ratio: 
\begin{equation}  
\label{hob_wallratio}
r_{n/m} = \frac{e_n}{e_m} 
 \end{equation} 
In figure \ref{figwall} these ratios are plotted in the case $n=3, m=2$ for different $\alpha$ as a function of the inverse power $1/p$ with thick lines.
 
In order to obtain an extrapolation for $p \rightarrow \infty$, we fit $r_{n/m}$ with an inverse polynomial $r^{\textrm{fit}}_{n/m}$, which is explicitly given by 
\begin{equation}  
\label{hob_wallratiofit}
r^{\textrm{fit}}_{n/m}(p) = c_0 + \frac{c_1}{p}+ \frac{c_2}{p^2} 
 \end{equation} 
by determining the coefficients $c_0, c_1, c_2$.

In figure \ref{figwall} these ratios are plotted using dotted lines. The limiting values for the infinite potential at $1/p \rightarrow 0 $ are simply given as $\lim_{ p \rightarrow \infty} r^{\textrm{fit}}_{n/m} = c_0$  and are plotted as crosses at $1/p = 0$.

There are no exact analytic solutions for eigenvalues and -functions of the fractional  Schr\"odinger equation with infinite potential well potential. 

\begin{figure}[t]
\begin{center}
\includegraphics[width=\textwidth]{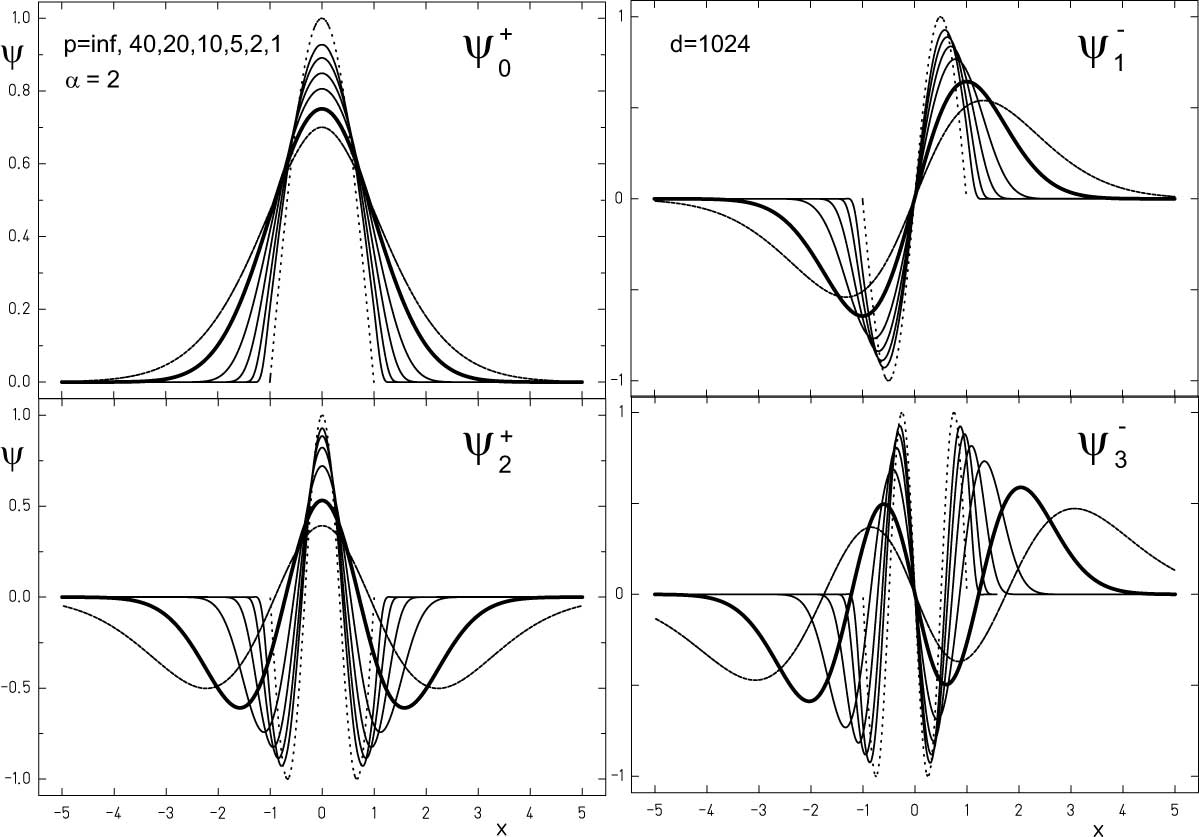}    
\caption{Eigenfunctions of the fractional harmonic oscillator for $\alpha=2$  and increasing power $p$ of potential $\frac{1}{2}|x|^p$. In the limit for $p \rightarrow \infty$  the eigenfunctions of the infinite potential well are given with pointed lines. 
}
\label{figwellpl2}
\end{center}
\end{figure}

As a check for the accuracy of the diagonalization procedure, in figure \ref{figwellpl2} we show the numerical results for the lowest four eigenfunctions for the classical case $\alpha=2$ within a range of $p$ starting from moderate $p=1$ (linear potential)  up to $p=40$. For the infinite potential well ($p\rightarrow \infty$) the classic solutions
$\sin(k x)$ and $\cos(k x)$ respectively are plotted with dotted lines to mark the ideal case. The shape (thick lines) of the calculated numerical solutions develop in a 
reasonable way towards the ideal case.    

Approximate solutions for the eigenvalues, based on the trigonometric $\sin$ and $\cos$ functions,  which ignore the nonlocality of the problem have been published e.g. by Laskin \cite{laskin}:
\begin{eqnarray}  
\label{hob_walle0}
e^{\textrm{class}}_n &=& (\frac{\pi}{2} n )^\alpha
\end{eqnarray} 
another approximation results from  the WKB-approximation
\begin{eqnarray}  
\label{hob_walleWKB}
e^{\textrm{WKB}}_n      &=& \lim_{p \rightarrow \infty} e^{\textrm{WKB}}(n,\alpha,p)  \\
&= & \left( \frac{\pi}{4}(n+\frac{1}{2}) \right)^\alpha
\end{eqnarray} 
In a recent work \cite{he12b}, calculating the energy expectation values  for the infinite potential well using the trigonometric $\sin$ and $\cos$ test functions, we derived another approximation given as: 
\begin{eqnarray}  
\label{hob_wallecs}
e^{\textrm{cs}}_n  &=& \frac{\sin(\frac{\alpha \pi}{2}) \Gamma(\alpha)}{2 \pi}  \times \\
&& (\frac{\pi^2 n^2 \,{_1}F_2((1-\frac{\alpha}{2};\frac{3}{2},2-\frac{\alpha}{2};(\frac{n \pi}{4})^2)}{\alpha-2} + 8 \sin^2(\frac{n \pi}{4})) \nonumber
\end{eqnarray} 
Based on this approximate spectra the corresponding energy ratios follow as:
\begin{eqnarray}  
\label{hob_rclass}
r^{\textrm{class}}_{n/m}      &=& 
 \left(
  \frac{n}{m}
 \right)^\alpha \\
\label{hob_rWKB}
r^{\textrm{WKB}}_{n/m}      &=& 
 \left(
  \frac{n+1/2}{m+1/2}
 \right)^\alpha \\
\label{hob_rcs}
r^{\textrm{cs}}_{n/m}      &=& 
  \frac{\pi^2 n^2 \,{_1}F_2(1-\frac{\alpha}{2};\frac{3}{2},2-\frac{\alpha}{2};(\frac{n \pi}{4})^2) + 8 (\alpha-2) \sin^2(\frac{n \pi}{4})}
  {\pi^2 m^2 \,{_1}F_2(1-\frac{\alpha}{2};\frac{3}{2},2-\frac{\alpha}{2};(\frac{m \pi}{4})^2) + 8 (\alpha-2) \sin^2(\frac{m \pi}{4})} \nonumber\\
  &&
\end{eqnarray} 
In figure \ref{figwall} all ratios are plotted for the case $r_{3/2}$ for different $\alpha$. Numerical results are compared with the above given in 
(\ref{hob_rclass}), (\ref{hob_rWKB})  and (\ref{hob_rcs}) approximations. 

The comparison of approximate methods for a computation of eigenvalues and derived ratios plotted is an additional indicator for the quality and accuracy  of the
calculated results.

The simple   
$k$ power law (\ref{hob_rclass}) (hollow circles) yields good results near $\alpha=2$ because this is the classical limit for the infinite potential well, but for $\alpha \ll 2$ the
 calculated ratio overestimates the numerical result. 
 
The WKB-results (dashed lines and stars)  are a good choice for moderate $p$-values, but for large $p$ they  increasingly deviate from numerical results. 
 
Finally, calculated approximations $r^{\textrm{cs}}_{3/2}$ (crosses)  obviously underestimate the numerical results. 

Summarizing the results we obtain a clear signal for the statement, that the simple  power law (\ref{hob_walle0}) with the assumption, that the eigenfunctions of
the fractional infinite potential well are simple trigonometric functions, fails for $\alpha \neq 2$, which is in agreement with previous findings \cite{he12b, luc13, duo14}
and proves wrong statements made by \cite{las10, guo06, don07, bay12a, bay12b}, which should be considered merely as approximations valid for $\alpha \approx 2$ only  \cite{jen, haw12}.
 
\section{Unexpected phenomena - Level crossings and a new parity symmetry}
\index{level crossings}
\begin{figure}[t]
\begin{center}
\includegraphics[width=\textwidth]{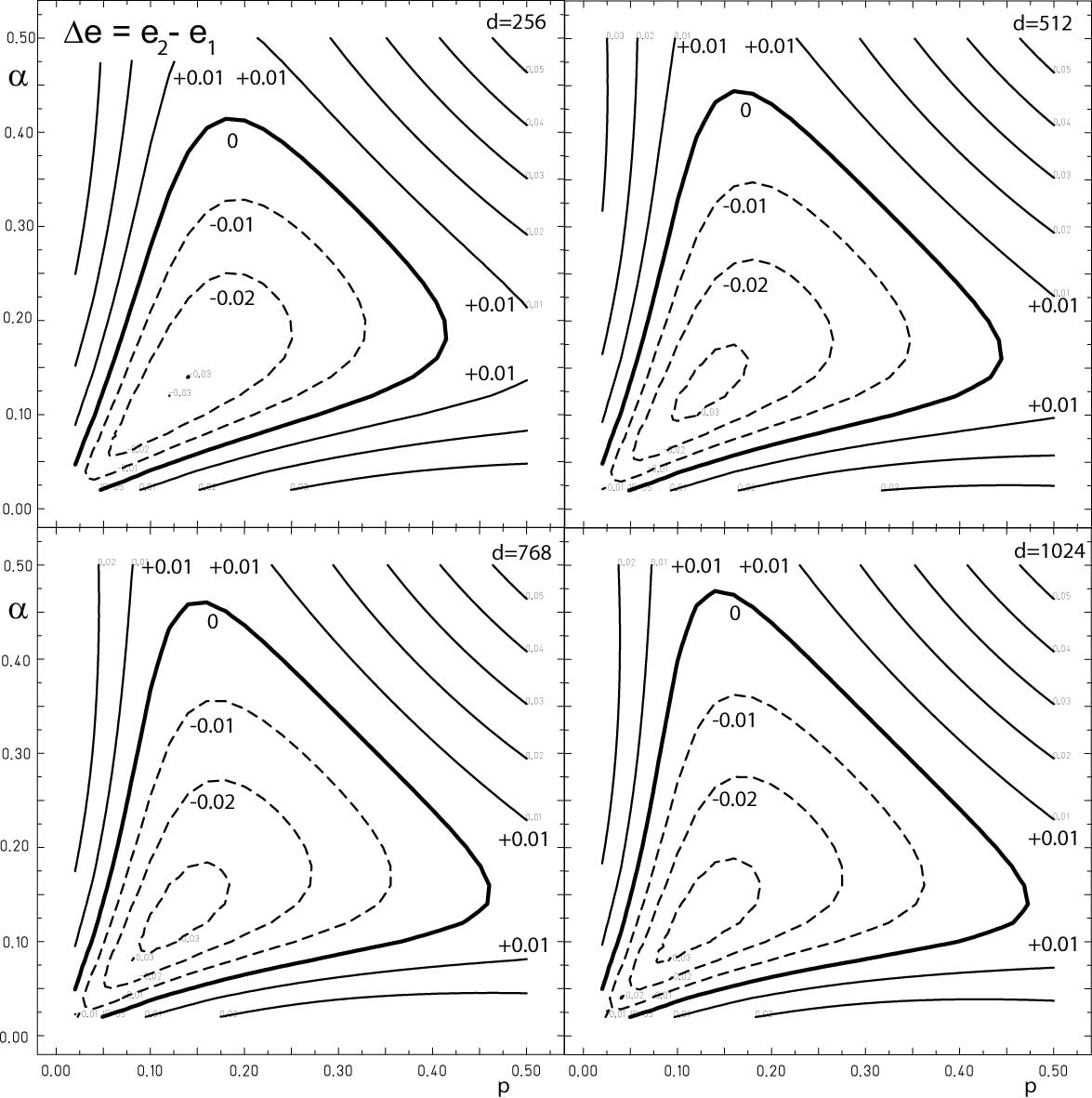}    
\caption{Contour plot of energy difference $\Delta e = e_2 - e_1$ of the fractional harmonic oscillator energies of the first eigenfunctions $\psi^+_2$ and $\psi^-_1$ in the range $0.02 \leq \alpha \leq 0.5$ and $0.02\leq p \leq 0.5$ for different matrix dimensions $d = 256, 512, 768, 1024$. Solid thin lines indicate $\Delta e \geq 0$, thick line indicates $\Delta e=0$, dashed line indicates negative values. For the
classical harmonic oscillator ($\alpha=p=2$) this difference is constant $\Delta e = e_{n+1} - e_n, \, \forall n$. With decreasing either $\alpha$ or $p$ this
difference is reduced. For both values small enough we obtain a negative energy difference for $n=1,3,5,...$, which implies a level crossing. Here, for the case $n=1$ the energy of the $2^+$-state falls below the energy of the $1^-$-state.
}
\label{figwallcross}
\end{center}
\end{figure}
From the  results presented so far, the matrix representation of  the fractional  Schr\"odinger equation has indeed proven to be  an appropriate approach to obtain 
high accuracy, reliable results for the eigenvalues and functions of the fractional harmonic oscillator.  Furthermore we observe at first unexpected phenomena, which
lead to a deeper understanding of the symmetries of the fractional extension of the standard harmonic oscillator.

In particular,  we observe two new phenomena, which are more or less  directly connected to symmetry: 

One is a level crossing for even-odd energy levels in the region $\alpha, p < 0.5$. 
\index{level crossing}

For the classical quantum mechanical harmonic oscillator ($\alpha=p=2$) and its level spectrum $e_n=n+1/2$ we have an equal spacing of energy levels $\Delta e = e_{n+1} - e_n=1$ or in other words  $\Delta e >0,\, \forall n$. Any deviation of the set of parameters c from the classical value pair $\{2,2\}$ results in a non-equidistant level spectrum, but in a first guess, we may assume, that $\Delta e >0,\,\forall n$ still holds for any pair of $\Delta e >0,\, \forall n$, which means, that the energy differences may
deviate from the equidistant, but the sequence of energy levels is determined by increasing the value of $n$ in the fractional case,  too.

This guess is wrong in the region   $\{\alpha < \sim 0.5, p < \sim 0.5\}$, the fractional harmonic oscillator spectra are not that simple, as we will demonstrate in the following. 

In figure~\ref{figwallcross} the energy difference $\Delta e = e_2 - e_1$ is plotted in the
region  $\{0.02 \leq \alpha \leq 0.5$, $0.02\leq p \leq 0.5\}$ for increasing matrix dimensions, so we can observe the stabilization of the phenomenon for increasing the
accuracy. For larger values of $\{\alpha, p\}$ the energy difference is positive, but the thick line shows the line of degeneracy $e_2 = e_1$,  which is the boundary for a region with $\Delta e < 0$. 

Since the physical origin of a degeneracy in a quantum-mechanical system is often the presence of an additional symmetry in the system. 
A typical example are the eigenvalues of a two dimensional rectangular membrane with sides $a,b$, where degeneracy occurs for the special case of a square $a=b$.    

The shape of the corresponding wave functions does not change drastically. We therefore observe a real level crossing, not an avoided one like
e.g. in non-adiabatic level crossing scenarios, see Zener \cite{zen32}, where characteristics are exchanged. 

Inside the bounded region condition  $\Delta e < 0$ holds. This behavior is specific to the Riesz definition of the fractional derivative. For the Caputo- and Riemann- based
definitions of the fractional derivative for the fractional quantum harmonic oscillator we observe an increasing dying out of energy levels  at least along the $\alpha=p$ line \cite{he13a} instead.

We consider this result as a first hint for an inherent additional symmetry for the fractional quantum oscillator.

\begin{figure}[t]
\begin{center}
\includegraphics[width=\textwidth]{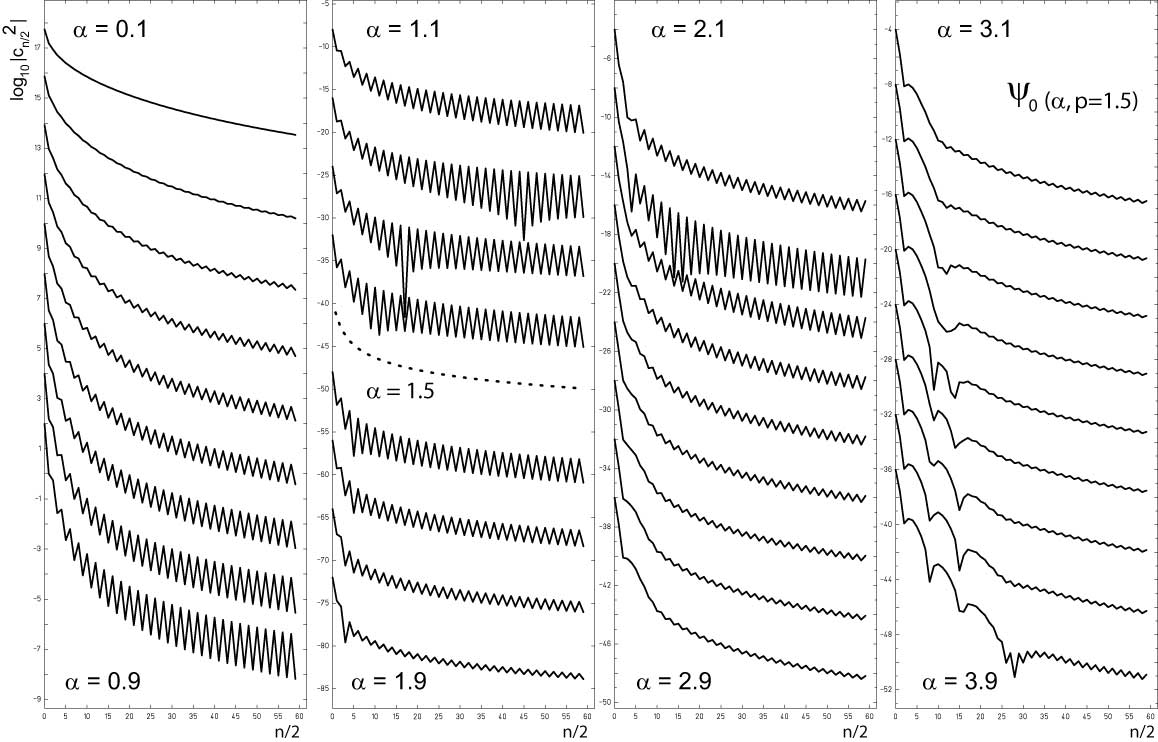}    
\caption{For $p = 1.5$ and different $\alpha$ the  absolute value of the amplitudes $|c_{0, 2n}|$  in the series expansion of harmonic oscillator basis functions for the eigenvector of the ground state $\Psi_0^+(x)$ is plotted. Since the ground state has even parity, only coefficients for even basis functions deviate from zero, while the odd coefficients vanish $|c_{0, 2n+1}| = 0$. The
parity property is conserved for the fractional case because of the specific parity conserving properties of the Riesz-weight. The presence and dominance of an additional symmetry is 
revealed near the higher symmetric case $\alpha=p$, where in addition every second even coefficient vanishes $|c_{0, 4n+2}|=0$ and in the wider region near $\alpha=p$ is
significantly suppressed.
}
\label{fig225coeff}
\end{center}
\end{figure}
%

There is another phenomenon, which is even more fundamental and will lead to more drastic simplifications for the  numerical treatment of the fractional kinetic energy
matrix elements.
\begin{eqnarray}  
\label{hob_patiy}
\Psi_m(x) &=& \sum_{n=0}^d  c_{m, n} \psi_n(x), \qquad 
\end{eqnarray}
The effort to solve the eigenproblem for given matrix-dimension $d=n_{max}$ is in general of order \cite{pan99}: 
\begin{eqnarray}  
\label{hob_order}
\mathcal{O}(d^3 + (d \log^2(d) log (b))  
\end{eqnarray}
$\mathcal{O}(z) $ being  the big O-notation for the asymptotic behavior for a function  \cite{bac94, lan09} and $b$ is a constant.
If an additional symmetry is present, this may be used to reduce the matrix-dimension for a given accuracy goal.
A classic example is parity conservation and 
a special property of the Riesz-weight is indeed parity conservation. 

As a consequence, for the fractional quantum oscillator based on the Riesz-definition of a fractional derivative 
parity is conserved and therefore the corresponding eigenfunctions carry the conserved quantum number parity, which means, that solutions are given by
\begin{eqnarray}  
\label{hob_patiy2}
\Psi^+_m(x) &=& \sum_{n=0}^{d} c^+_{m, 2 n} \psi_{2 n}(x), \qquad\qquad  m \, \textrm{even}\\
\Psi^-_m(x) &=& \sum_{n=0}^{d} c^-_{m, 2 n+1} \psi_{2 n+1}(x), \qquad m \, \textrm{odd} 
\end{eqnarray} 
This may be written in compact form:
\begin{eqnarray}  
\label{hob_patiy2c}
\Psi^{mod(m,2) (-1)}_m(x) &=& \sum_{n=0}^{d} c_{m, 2 n + mod(m,2)}^{mod(m,2) (-1)} \psi_{2 n+  mod(m,2) }(x),       \, m \in N_0\nonumber\\
&&
\end{eqnarray} 
$mod(i,j)$ is the $i$ modulo $j$ function, which reduces the matrix dimension by a factor 2
and therefore the effort for a solution of the eigenproblem is  reduced  according (\ref{hob_order}) by a factor 4.

In figure \ref{fig225coeff} the absolute value of the coefficients $c_{0, 2 n}$ for the ground state eigenfunction $\Psi_0^+(x)$ for a given fixed power $p=1.5$ are plotted for increasing $\alpha$.  The presence and dominance of an additional symmetry is 
revealed near the high symmetric case $\alpha=p$, where every second even coefficient vanishes $|c_{0, 4n+2}|=0$ and is significantly suppressed in the vicinity, which leads to the observable oscillatory behavior of the coefficients.

In fractional quantum mechanics the pairs $\{\alpha,p\}$ and $\{p,\alpha\}$ lead to the same eigenfunctions and eigenvalues because the equivalence of
coordinate and momentum representation or from a more mathematical point of view, the Fourier-transform of kinetic and potential energy operator respectively. 
For the special case $\alpha = p$ the Fourier-transform of  the kinetic and potential part of matrix elements cancel each other out  exactly in the case  $m=0$ and $n=4k+2$.

Hence for $\alpha = p$ we obtain 4 cases:
\begin{eqnarray}  
\label{hob_patiy4}
\Psi^{+1}_m(x) &=& \sum_{n=0}^{d} c^{+1}_{m, 4 n } \psi_{4 n}(x),        \qquad\quad \qquad m=0,4,8,12,...\\
\Psi^{+i}_m(x)  &=& \sum_{n=0}^{d} c^{+i}_{m, 4 n+1} \psi_{4 n+1}(x),  \qquad\quad m = 1,5,9,13, ...\\ 
\Psi^{-1}_m(x)  &=& \sum_{n=0}^{d} c^{-1}_{m, 4 n+2} \psi_{4 n+2}(x),   \qquad\quad  m = 2,6,10,14,...\\
\Psi^{-i}_m(x)   &=& \sum_{n=0}^{d} c^{-i}_{m, 4 n+3} \psi_{4 n+3}(x),   \qquad\quad  m = 3,7,11,15,...
\end{eqnarray} 
which may be written in compact form:
\begin{eqnarray}  
\label{hob_patiy4}
\Psi^{mod(m,4)  i}_m(x) &=& \sum_{n=0}^{d} c_{m, 4 n + mod(m,4)}^{mod(m,4) i} \psi_{4 n+  mod(m,4) }(x),       \, m \in N_0 \nonumber \\
&&
\end{eqnarray} 
where $i = \sqrt{-1}$ .
Hence for $\alpha = p$ the effort to solve the corresponding eigenproblem is reduced in total by a factor 16. 

The corresponding symmetry may be interpreted as an extended parity invariance of the problem which is extended to the complex case. While the standard parity operator
 $\Pi$ generates in coordinate space a rotation of $x$ by $\pi$ or $180\degree$, which yields the sequence $\{-1,1\}$:
\begin{eqnarray}  
\label{hob_patiy4ex}
\Pi^2(180\degree) &=& 1
\end{eqnarray} 
we now require
\begin{eqnarray}  
\label{hob_patiy44ex}
\Pi^4(90\degree) &=& 1
\end{eqnarray} 
which in coordinate space corresponds to a rotation of $x$ by  $\pi/2$ or $90\degree$, which yields a sequence $\{i, -1, -i, 1\}$ of coordinate factors. The requirement of invariance of the
wave equation under this quartic parity operation is fulfilled by omitting the corresponding coefficients in the series expansion in terms of the harmonic oscillator
basis functions.

Once this symmetry is revealed, we may use it to postulate a direct correlation between kinetic and potential energy matrix elements.

Indeed we obtain an unexpected simplification for the kinetic energy matrix elements:
\begin{numcases}
{\braket{m|T(\alpha)|n} =} 
\label{c255plus}
+\braket{m|V(\alpha)|n}         &  $|(m-n)/2| \, \,\textrm{even}$\\
 -\braket{m|V(\alpha)|n} &   $|(m-n)/2| \, \,\textrm{odd}$
 \end{numcases}
 which is the fractional matrix equivalent of  (\ref{hob_x22}) and (\ref{hob_p22}). Since $\braket{m|V(\alpha)|n}$ is given by (\ref{hob_potreal}), we actually
 obtain an unexpectedly short closed formula for the kinetic energy matrix elements. 
 
 Let us summarize these amazing results, collected in this investigation:
 \begin{itemize}
 \item The matrix representation of the fractional Schr\"odinger equation is an ideal starting point to solve eigenproblems in non-relativistic fractional quantum mechanics. 
 Especially the separate treatment of non-local kinetic energy and classical potential  avoids a repeated calculation of complex non-local matrix elements for different potentials.
 Once the kinetic energy matrix elements are known, they may be used in many different scenarios. 
 
 \item The choice of the harmonic oscillator basis allows for a correct treatment of weakly singular kernels as well as long range interactions. 
 
 \item  Furthermore the matrix representation
 of the fractional harmonic oscillator, here using the Riesz definition of a fractional derivative,  is the basic key for an understanding of inherent symmetries, which otherwise could not be investigated  adequately.

Thus some unexpected features were detected and described: 

Light houses in the error land map, characterized by a significant increase of precision for even integer positions $\{ \alpha = 2n, p = 2m  \}$ $n,m \in N_0$. 

Level crossings for level pairs $e_{2 n}- e_{2 n-1}, \,\forall n$. 

The conservation of particle-wave dualism perpetuated to fractional quantum mechanics using the Riesz definition of a fractional derivative, extends the parity symmetry to the fractional case, leading to a generalized
fractional parity symmetry extended to the complex plane.
In coordinate space it corresponds to a rotation of $x$ by  $\pi/2$ or $90\degree$, which yields a sequence $\{i x, -x, -i x, x\}$ of function arguments. The requirement
of invariance of the wave functions under this extended parity operation lead to an unexpected simplification and was the key indication for the significant 
simplification of the kinetic energy matrix element calculation. 
\index{fractional!parity}  
\index{parity}  
 
 \item  The approach
 culminated in an unexpectedly simple closed explicit representation of the fractional kinetic energy matrix elements for the Riesz kernel.
 
 \item  Furthermore the presented approach opens a new vista on the possible practical interpretations of a fractional derivative in terms of a specific superposition of Glauber-states.  
\end{itemize}
 Some open tasks  remain:
 
 The step from integer to real power for the oscillator potential, which means, the step from (\ref{hob_potint}) to (\ref{hob_potreal}) still needs a thorough investigation. 
 In addition the correspondence between kinetic and potential matrix elements heuristically introduced in (\ref{c255plus}) needs a formal proof as well.
 
 Considering all in all, the matrix representation is a serious alternative to the standard formulation of fractional quantum mechanics in terms of fractional wave equations and recommends itself to be 
 a strong formal basis for further investigations and applications e.g. a fractional perturbation theory.

\section{Acknowledgments}
We thank A. Friedrich  for useful discussions.

%

\end{document}